\documentclass[amsmath,amssymb,aps, pra,preprint, superscriptaddress, nofootinbib]{revtex4-1}

\usepackage{graphicx}
\usepackage{comment}
\usepackage{cancel}
\usepackage[normalem]{ulem}
\usepackage{tabularx}
\usepackage{makecell}
\usepackage{multirow}
\usepackage{amsmath,amssymb}
\usepackage{array}
\newcolumntype{P}[1]{>{\centering\arraybackslash}p{#1}}
\usepackage{nicematrix}

\usepackage{dcolumn,soul,bbold}
\usepackage{orcidlink, float}

\newcommand{\comlm}[1]{\textcolor{black}{#1}}

\newcommand{\comjt}[1]{\textcolor{black}{#1}}
\newcommand{\comcb}[1]{\textcolor{black}{#1}}

\begin{document}


\title{Bound States and Resonance Analysis of One-Dimensional Relativistic Parity-Symmetric Two Point Interactions}


\author{Carlos A. Bonin \orcidlink{0000-0002-9005-3100}\, }
\email[]{carlosbonin@gmail.com}
\affiliation{{Independent Scholar, Brazil}}
\author{Manuel Gadella \orcidlink{0000-0001-8860-990X}\, }
\email[]{manuelgadella1@gmail.com}
\affiliation{Departamento de F\'{\i}sica Te\'orica, At\'omica Y \'Optica and IMUVA, Universidad de Valladolid, 47011, Valladolid, Spain}
\author{Jos\'e T. Lunardi \orcidlink{0000-0001-7058-9592}\, }
\email[]{jttlunardi@uepg.br}
\affiliation{Department of Mathematics \& Statistics, State University of Ponta Grossa, Avenida Carlos Cavalcanti 4748, Cep 84030-900, Ponta Grossa, PR, Brazil}

\author{Luiz A. Manzoni \orcidlink{0000-0002-0035-9529}\, }
\email[]{manzoni@cord.edu}
\affiliation{Department of Physics, Concordia
College, 901 8th St. S., Moorhead, MN 56562, USA}


\date{\today}

\begin{abstract}

We consider the one-dimensional Dirac equation with the most general relativistic contact interaction supported on two points symmetrically located with respect to the origin. In order to determine the shape of the interaction, we use a distributional method, which in the present case is equivalent to the standard method of defining contact interactions by self-adjoint extensions of symmetric operators. The interaction on each of these two points depends on four parameters, each one having a clear physical meaning.  We are interested in the scattering and confining properties of this model. We focus our attention on even or odd interactions under parity transformations and investigate the existence of critical and supercritical states, bound states, confinement and scattering resonances for some particular interactions of special interest.

\end{abstract}


\maketitle







\section{\label{intro}Introduction}

One-dimensional point interactions (PI) in quantum mechanics have attracted significant attention, both for their intrinsic mathematical and theoretical appeal and the flexibility afforded by its 4-parameter characterization  \cite{AGH04, Seb86, Car93, Kur96, CNP97}, which enables modeling of a wide range of short-range interaction systems. From a mathematical and theoretical point of view, the properties of these interactions have been extensively studied using methods such as regularization \cite{Seb86, Er17, Fas19, Jac91, RTa96}, renormalization \cite{TurgutEroncel2014ERGPointInteractions}, self-adjoint extensions (SAE) \cite{AGH04, Fas18, CNP99}, and distribution theory \cite{Kur96, LMM13, CLM14, Lan15, DJP16, CLM19}. PIs also serve as idealized models in numerous physical contexts. Notable examples include their role in exactly solvable many-body systems such as the Lieb-Liniger model \cite{LLi63, Lie63} and the Tonks-Girardeau gas \cite{Ton36, Gir60,KWW04, Settino2021TonksGirardeau}, as well as in investigations of the Casimir effect \cite{Silva2016DynamicalCasimir, Bordag2017CasimirDiracLattices},  Y-junctions \cite{Tkachenko2012YJunctionSwitch}, and other quantum systems with localized interactions. For an overview of recent developments in both the theoretical treatment and practical applications of one-dimensional contact interactions, see \cite{GML21}. 

The non-relativistic scattering by two general point interactions has been analyzed in connection with tunneling times \cite{LLM16}. The resonant structure of such systems, which is of particular relevance for junction heterostructures \cite{Ba81,vR83,Ga07}, has been further investigated in \cite{KNT16,KNT17} (see also \cite{Boy08}). In \cite{FGN24}, the authors examined the one-dimensional Hamiltonian $-d^2/dx^2$ modified by two non-local $\delta'$ interactions, employing a renormalization approach to the coupling constant. Relativistic point interactions, on the other hand, may provide corrections to non-relativistic models and are relevant for describing impurities in quasi-one-dimensional graphene nanoribbons \cite{BFe06}. These systems have been investigated by many authors, \textit{e.g.}, \cite{SMa81, MSt87, DMa89, Roy93, CAZ03, ACG11, ASh15, HTu22, BLM24}. Similarly to the non-relativistic case, they also constitute a four-parameter family of interactions \cite{FGr87, BDa94}.  In particular, the scattering of a relativistic particle by two point interactions was originally {analyzed} in \cite{Roy93a}, where the author considered two $\delta$ barriers as given by the appropriate limit of rectangular barriers and obtained resonant tunneling at all energies for certain values of the interaction strength, a result later disproved in \cite{Yan02}. However, a systematic study of relativistic resonant scattering for an arrangement of two point barriers seems to be lacking in the literature and will be addressed in this work, particularly in the case of arrangements having well-defined parity.

At this point, it may be convenient to recall that {\it scattering resonances} are characterized by diverse but equivalent conditions, such as i) pairs of poles of the analytic continuation of the $S$-matrix (either on the energy or momentum representation), or ii) singularities of the transmission coefficient, or, iii) zeroes of the Jost function or, iv) the so called {\it purely outgoing boundary condition}, in which the coefficient of the incoming wave function is taken equal to zero for certain (complex) values of the energy (equivalently, of the momentum). Pairs of values of the energy  characterizing scattering resonances are always complex and the members of each pair are complex conjugate of each other. See \cite{New82,Nus72,KKH89,Exn85,GPr11,AGH01,Boh94}. We shall discuss the emergence of scattering resonances in our model. 

In order to carry out a thorough analysis of the two-point relativistic interaction, we employ the distributional approach (DA) to contact interactions developed in \cite{LMM13, CLM14, CLM19, BLM24}. This framework provides an explicit formulation of the interaction terms and expresses them directly in terms of the physical fields \cite{BLM24} (see also \cite{BHT23}). Such a formulation is particularly well suited for  symmetry analysis.

This paper is organized as follows: In Section II, we recall the most general form of the two-point relativistic contact interactions and their transformation under parity transformations. In Section III, the form of the wave function and the transmission matrix are introduced and the single point interaction limit is discussed. Section IV is devoted to a general analysis of critical, supercritical and bound states, and scattering resonances, leaving a detailed study of special cases of particular interest for Section V. The paper closes with some Concluding Remarks plus two Appendices.  In the first appendix, we establish the relations between the four physical parameters that fix the distributional form of each of the contact interactions, with those parameters that fix the matching conditions of the wave functions on each point supporting the interactions. Each of these matching conditions give a self-adjoint determination of the Hamiltonian under study. In the second appendix, we give the non-relativistic limit of some selected  Dirac equations derived from some of these self-adjoint determinations of the Hamiltonian.


\section{Analysis of Symmetry under Parity Transformation}

The one-dimensional stationary Dirac equation for a particle of mass $m$ and energy $E$ scattered by a double-barrier of singular interactions can be written as a distributional differential equation of the form (throughout this paper we adopt natural units, $\hbar = c = 1$) \cite{CLM14, BLM24}
\begin{equation} \label{dir}
\left(i \gamma^1 \partial_1 -  m\mathbb{1}  +\gamma^0E \right) \psi (x) = D[\psi](x) \; ,
\end{equation}
where $\gamma^0 = \sigma_3$ and $\gamma^1 = i\sigma_2$, with $\sigma_i$ indicating the Pauli matrices. The distributional interaction $D[\psi](x)$ represents the most general distribution associated with the singular interaction. 

Here we consider a double barrier of singular interactions placed at points $x=\pm \ell/2$. Then, following \cite{BLM24}, { $D[\psi](x)$ is given by ($\mu\in\{0,1\}$)}
\begin{eqnarray}
D[\textcolor{black}{\psi}](x)&=& \left( B^{(1)} \mathbb{1} + A^{(1)}_{\mu}\gamma^{\mu} + iW^{(1)}\gamma^5 \right) \delta(x+\tfrac{\ell}{2}) \,\frac{\psi (-\tfrac{\ell}{2}^+)+\psi 
(-\tfrac{\ell}{2}^-) }{2} \nonumber \\  \label{d} \\
&+& \left( B^{(2)} \mathbb{1} + A^{(2)}_{\mu}\gamma^{\mu} + iW^{(2)}\gamma^5 \right) \delta(x-\tfrac{\ell}{2}) \,\frac{\psi (\tfrac{\ell}{2}^+ )+\psi ( \tfrac{\ell}{2}^- )}{2} \; .\nonumber
\end{eqnarray}
In this expression{, $\gamma^5 = \gamma^0\gamma^1 = \sigma_1$} and $\psi (x^{\pm}) \equiv \lim_{\varepsilon \to 0^{\textcolor{black}{+}}} \psi (x \pm \varepsilon)$ stand for the {one-sided} limits of the spinor distribution $\psi (x)$ (which, if they exist, can be defined even at singular points \cite{Loj57, VEs07}). The so-called physical parameters $B^{(j)}$, $A_{\mu}^{(j)}$, $W^{(j)}$, with {$j\in\{1,2\}$}, are the strengths of the scalar, vector, and pseudoscalar point interactions\footnote{\comcb{This means the following: if $a_j$ is an arbitrary Lorentz vector, $u$ is a unit timelike vector, and $\epsilon_{\mu\nu}$ is a component of the 2-dimensional Levi-Civita symbol, then $B^{(j)}\delta\left(\epsilon_{\alpha\beta}u^\alpha\left(x^\beta-a^\beta_j\right)\right)$  transforms between compatible inertial frames as a Lorentz scalar field, $A_{\mu}^{(j)}\delta\left(\epsilon_{\alpha\beta}u^\alpha\left(x^\beta-a^\beta_j\right)\right)$  as a component of a Lorentz vector field, and $W^{(j)}\delta\left(\epsilon_{\alpha\beta}u^\alpha\left(x^\beta-a^\beta_j\right)\right)$ as a Lorentz pseudoscalar field,  as shown by some of us for the case $a_j\equiv 0$ in \cite{BLM24}.}} located at $x_1 = -\ell/2$ and $x_2= +\ell/2$, respectively. 

The analysis of the properties of the interaction under space reflections, as well as the physical interpretation of the interaction parameters, is more readily done in terms of the physical parameters in (\ref{d}), thus this will be the main form adopted in this section. However, the bound states and resonances{, to be calculated in the following sections,} are often expressed more compactly in terms of the familiar $\Lambda$-matrix parameters used in SAE and establishing the boundary conditions for the spinor $\psi(x)$ across the singular points as 
\begin{equation}\label{bc}
\psi(x_j^+)=\mathbb{\Lambda}_j\psi(x_j^-)\; ,
\end{equation}
with $\mathbb{\Lambda}_j$, {$j\in\{1,2\}$}, given by the well-known expression \cite{BDa94, CLM14}
\begin{equation}
\label{lbda}
\mathbb{\Lambda}_j ={\mathrm{e}}^{i\varphi_j}\left(
\begin{array}{cc}
a_j&ib_j\\
-ic_j&d_j
\end{array} \right), \quad a_jd_j-b_jc_j=1,
\end{equation}
where $\varphi_j\in[0,\pi)$ and $a_j,b_j,c_j,d_j \in \mathbb{R}$ are (dimensionless) constants. The relationships between the $\mathbb{\Lambda}$-matrix parameters in (\ref{lbda}) and the physical parameters in (\ref{d}) were explicitly obtained in \cite{BLM24} (see also \cite{HTu22, BHT23}
) and, for convenience, are reproduced in Appendix {\ref{app strengths}}. \comjt{The point interaction is said \emph{permeable} (or \emph{penetrable}) at $x_j$ ($j=1,2$) if all the $\mathbb{\Lambda}$-matrix parameters are finite; otherwise, it is said to be \emph{impermeable} (or \emph{impenetrable}) at $x_j$. See the Appendix  {\ref{app strengths}} for the conditions of permeability of a point interaction in terms of the physical parameters.}  

Since we are interested in interactions with well-defined parity, let us analyze how the interaction term in (\ref{dir}) and (\ref{d}) transforms under a space-reflection {transformation} $\cal P$. The transformation properties of the distributional spinor $\psi (x)$ under $\cal P$ are obtained much in the same way as for the usual (non-distributional) Dirac equation \cite{BDr64, GMP19}, and it can be shown that in a particular (laboratory) reference frame we have \cite{BLM24}
\begin{eqnarray}\label{parityT}
&&\mathcal P:\, \left(t,x \right)\mapsto \left(t,x^{\prime}\right)=\left(t,-x \right) \; ,
\\
\label{psiparity}
&&\mathcal P:\, \psi(x)\mapsto  \psi^P\left(x^{\prime}\right)=\gamma^0\psi\left(x\right)=\gamma^0\psi\left(-x^{\prime}\right) \; , \\ 
\label{Vparity}
&&\mathcal P:\, D[\psi ](x)\mapsto D^P[\psi^P]\left(x^{\prime}\right)=\gamma^0  D[\psi]\left(-x^{\prime}\right) \; .
\end{eqnarray} 

Then, noticing that under space reflection $\delta (x\pm \ell/2) \to \delta (x^{\prime} \mp \ell /2)$, $\psi(\textcolor{black}{+}\ell^\pm /2) \to \psi^{P\, \textcolor{black}{}}(\textcolor{black}{-}\ell^\mp /2)$ and $\psi^{}(\textcolor{black}{-}\ell^\pm /2) \to \psi^{P\, \textcolor{black}{}}(\textcolor{black}{+}\ell^\mp /2)$, we obtain
\begin{eqnarray}
D^P[\psi^P](x^{\prime})&=& \left( B^{(1)} \mathbb{1} + A^{(1)}_0 \gamma^0-A^{(1)}_1\gamma^1 - iW^{(1)}\gamma^5 \right) \delta(x^{\prime}-\ell/2) \,\frac{\psi^{P}\left(\tfrac{\ell^+}{2}\right)+\psi^{P}\left(\tfrac{\ell^-}{2}\right)}{2} \nonumber \\ \label{dpar} \\
&+& \left( B^{(2)} \mathbb{1} + A^{(2)}_0 \gamma^0-A^{(2)}_1\gamma^1 - iW^{(2)}\gamma^5 \right) \delta(x^{\prime}+\ell/2) \,\frac{\psi^{P}\left(-\tfrac{\ell^+}{2}\right)+\psi^{P} \left(-\tfrac{\ell^-}{2}\right)}{2} \; \,.\nonumber
\end{eqnarray}
This expression is to be compared with (\ref{d}) to determine the specific symmetry of the {arrangement of interactions}, as below.

\noindent
\begin{itemize}
\item[--]\underline{\textit{Even {arrangement}}}

For an interaction to be even under space reflection, the interaction distribution must invariant, \textit{i.e.}, it must have the same formal expression after the transformation:
\begin{equation}
D^P[\psi^P](x^{\prime})=D[\psi^P](x^{\prime}) \; ,
\end{equation}
which, comparing (\ref{d}) and (\ref{dpar}), can be satisfied only if 
\begin{eqnarray}
&&B^{(2)} = B^{(1)}\equiv B \; , \label{Beven} \\
&&A_0^{(2)}=A_0^{(1)} \equiv A_0 \; , \\
&&- A_1^{(2)} = A_1^{(1)}\equiv A_1 \; ,\\
&&-W^{(2)}=W^{(1)} \equiv W \; . \label{Weven}
\end{eqnarray}

In terms of the $\Lambda$-matrix parameters, indicating by $\mathbb{\Lambda}_1$ the matrix at $-\ell/2$ and by $\mathbb{\Lambda}_2$ the matrix at $+\ell/2$, the {condition above} for even interaction can be written, using (\ref{ap})-(\ref{dp}){,} as
\begin{equation}\label{Leven}
\mathbb{\Lambda}_1^{{\mathrm{(e)}}}= {e}^{i\varphi}\left(
\begin{array}{cc}
a&ib\\
-ic&d
\end{array}
\right), \quad
\mathbb{\Lambda}_2^{{\mathrm{(e)}}}= {e}^{-i\varphi}\left(
\begin{array}{cc}
d&ib\\
-ic&a
\end{array}
\right),
\end{equation}
where the superscript ``(e)" stands for {\textit{even}. Notice that the \emph{individual} matrices $\mathbb{\Lambda}_i^{\mathrm{(e)}}$ do not necessarily possess any symmetry -- what is required is that their  \emph{arrangement} be symmetrically even \cite{BLM24}. }

\item[--] \underline{\textit{Odd {Arrangements}}}

For an interaction to be odd it must be such that
\begin{equation}
D^P[\psi^P](x^{\prime})=- D[\psi^P](x^{\prime}) \; ,
\end{equation}
which can be satisfied only if 
\begin{eqnarray}
&&B^{(2)} = -B^{(1)}\equiv -B \; ,\label{Bodd}\\
&&A_0^{(2)}=-A_0^{(1)} \equiv -A_0 \; ,\\
&&\textcolor{black}{A_1^{(2)} }= A_1^{(1)}\equiv A_1 \; , \\
&&\textcolor{black}{W^{(2)}}=W^{(1)} \equiv W \; ,\label{Wodd}
\end{eqnarray}
and the corresponding odd $\mathbb{\Lambda}$-matrices giving the {boundary conditions} at $x = \pm \ell/2$ again follow from (\ref{ap})-(\ref{dp}) {and} are given by
\begin{equation}\label{Lodd}
\mathbb{\Lambda}_1^{{\mathrm{(o)}}}= {e}^{i\varphi}\left(
\begin{array}{cc}
a&ib\\
-ic&d
\end{array}
\right), \quad
\mathbb{\Lambda}_2^{{\mathrm{(o)}}}= {e}^{i\varphi}\left(
\begin{array}{cc}
a&-ib\\
ic&d
\end{array}
\right).
\end{equation}
Again, the \emph{individual} matrices $\mathbb{\Lambda}_i^{\mathrm{(o)}}$ do not necessarily characterize an odd interaction, only their arrangement is odd \cite{BLM24}.

\end{itemize}

\section{General Eigenstates and 1-point Limits}
\label{solution}

Let us now consider the solution of the Dirac equation with two point interactions, obtaining {expressions for the transmission and reflection amplitudes}. The double barrier divides the space into three distinct regions{, in each of which} the Dirac particle moves freely. Thus, we can write the solution as
\begin{equation} \label{DirReg}
\psi(x) = \left\{
\begin{array}{l}
\psi_1(x)\, , \quad {\rm if}\;\; x<-\tfrac{\ell}{2};   \\
\psi_2(x)\, , \quad {\rm if}\;\;  -\tfrac{\ell}{2}<x<\tfrac{\ell}{2};   \\
\psi_3(x)\, , \quad {\rm if} \;\; \tfrac{\ell}{2}<x, 
\end{array}
\right.
\end{equation}
and, for $j\in \{1,2,3\}$, we have \footnote{{Notice that, according to the distributional approach, outside the support of the interaction, the distributional Dirac equation coincides with the usual Dirac equation.}}\footnote{\comcb{For the non-relativistic interpretation of the relativistic point interactions, it is often also convenient to rewrite the solution \eqref{psi} in the alternative form $\psi_j(x)=(u_j(x),v_j(x))^T$ – see Appendix \ref{app NR}.}}  
\begin{eqnarray}
\psi_j(x) &=&F_j e^{ikx}u_E(k) +G_j e^{-ikx}u_E(-k) \nonumber \\ \label{psi} \\
&\equiv & \mathbb{P}(k,x)
\left(
\begin{array}{c}
  F_j   \\
 G_j    
\end{array}
\right) \; , \nonumber
\end{eqnarray}
where $k\equiv \sqrt{E^{2}-m^{2}}$, $u_E(k)$ is the free Dirac spinor in momentum space, $u_E(k)= \left( 1, k/(E+m) \right)^T$ ($T$ here indicates the transpose), and we introduced the matrix $\mathbb{P}(k,x)$ defined by
\begin{equation}\label{m}
\mathbb{P}\left({k},x\right)=
\left(
\begin{array}{ccc}
e^{i k x}&&e^{-ik x}\\
\frac{k}{E+m}e^{ik x}&&-\frac{k}{E+m}  e^{-ik x}
\end{array}
\right)\, .
\end{equation}

The coefficients $F_j$, $G_j$ of the wave function outside the interaction region can be connected via the transfer matrix \textcolor{black}{$\mathbb{M}$} \cite{SMB12}, defined here as
\begin{equation} \label{transfer}
\left(
\begin{array}{c}
  F_3  \\
  G_3   \
\end{array}
\right)
= \mathbb{M}(\mathbb{\Lambda}_1,\mathbb{\Lambda}_2, k, x_1,x_2) 
\left(
\begin{array}{c}
  F_1  \\
  G_1  
\end{array}
\right)\; ,
\end{equation}
where
\begin{equation} \label{Matrix M}
\mathbb{M}(\mathbb{\Lambda}_1,\mathbb{\Lambda}_2, k, x_1,x_2) = \left[ \mathbb{P}^{-1}(k,x_2)\mathbb{\Lambda}_2 \mathbb{P}(k, x_2)\right] \left[\mathbb{P}^{-1}(k,x_1)\mathbb{\Lambda}_1 \mathbb{P}(k, x_1)\right] \; .
\end{equation}

%
%

\subsection{The limit $\ell\to 0^+$}

{We now} analyze the limit in which both singular interactions converge to a single point. To this purpose, it is convenient to define a connection matrix, $\mathbb{\Gamma}$, establishing a relationship between the wave function on both sides of the interaction as
\begin{equation}
\psi(x_2^+)=\mathbb{\Gamma} (\mathbb{\Lambda}_1, \mathbb{\Lambda}_2, k , x_1,x_2)\psi (x_1^-)\, ,
\end{equation}
that is,
\begin{equation}\label{Gamma}
\mathbb{\Gamma}\left(\mathbb{\Lambda}_1,\mathbb{\Lambda}_2,k,x_1,x_2\right) \equiv \mathbb{\Lambda}_2 \mathbb{P} \left( k,x_2\right) \mathbb{P}^{-1}\left(k,x_1\right)\mathbb{\Lambda}_1 = \textcolor{black}{\mathbb{P} \left( k,x_2\right) \mathbb{M} \,\mathbb{P}^{-1} \left( k,x_1\right)}\; .
\end{equation}

{From (\ref{Gamma}), we see that} $\lim_{x_2\to x_1}\mathbb{\Gamma} (\mathbb{\Lambda}_1, \mathbb{\Lambda}_2, k , x_1,x_2) = \mathbb{\Lambda}_2\mathbb{\Lambda}_1$, \comjt{since  $\lim_{x_2\to x_1}\mathbb{P} \left( k,x_2\right)=\mathbb{P}\left(k,x_1\right)$ in the middle of the above equation. Thus, the} composition law in this limit is given by the $U(1)\times SL(2,\mathbb{R})$ group structure, the same as in the non-relativistic case \cite{LLM16, GMM16} .  

From (\ref{Leven}) and (\ref{Lodd}), it follows that the single-barrier limit for even and odd arrangements of double barriers are, respectively, 
\begin{equation} \label{Lto0E}
\lim_{\ell \to 0^+} \Gamma^{(\text{e})} \!= \left(\!\!\!
\begin{array}{cc}
 1+2b c &  i\, \textcolor{black}{2bd} \\
 -i\, 2ac & 1+2b c \\
\end{array}
\!\!\!\right) ;
\end{equation}
and
\begin{equation} \label{Lto0O}
\lim_{{\ell \rightarrow 0^+}} \Gamma^{(\text{o})} \! =e^{2i\varphi} \left(\!\!\!
\textcolor{black}{\begin{array}{cc}
 a^2-bc & i b(a-d) \\
 i c(a-d)& d^2-bc  \\
\end{array}}
\!\!\!\right).
\end{equation}

The single-barrier limit of an even arrangement of interactions, eq. (\ref{Lto0E}), is an even single-point interaction \cite{CLM14, BLM24}, thus preserving the symmetry of the underlying double interactions, although the limit ${\ell\to 0^+}$ for odd arrangements of two singular barriers does not necessarily preserve the symmetry. In order to correspond to a single odd interaction, the off-diagonal elements of the $\Lambda$-matrix must be zero \cite{BLM24}; from (\ref{Lto0O}) this will occur if, and only if, $a=d$ \emph{or} $b=c=0$ (in terms of the physical strengths (\ref{Bpar})-(\ref{W}), this condition reads $W=0$ \emph{or} $B=A_0=0$, respectively). In the first case ($a=d$) neither of the two point interactions necessarily has a defined parity, the single point limit is $\lim_{{\ell \rightarrow 0^+}} \Gamma^{(\text{o})} = e^{i 2\varphi}\, \mathbb{1}$, which corresponds to an effective ``singular gauge field" interaction \cite{BLM24} (with the phase $2\varphi$ depending on all the potentially non-vanishing strengths $A_1$, $B$, and $A_0$ of the original arrangement of two point interactions).  The second case ($b=c=0$) corresponds to an odd arrangement of two general odd point interactions. The single point limit is
$$
\lim_{{\ell \rightarrow 0^+}}\Gamma^{(\text{o})} = e^{i 2\varphi } \left(
\begin{array}{cc}
 a^2 & 0 \\
 0& \frac{1}{a^2}  \\
\end{array}
\right)\,,
$$
which corresponds to a mixture of a pseudoscalar and a magnetic point fields (both depending on the parameters $W$ and $A_1$ of the original two point interactions). Finally, from (\ref{Lto0E}) and (\ref{Lto0O}) it is straightforward to observe that both even arrangements of two odd interactions and odd arrangements of two even point interactions converge to the free case when ${\ell\to 0^+}$.

In summary, in the limit ${\ell \rightarrow 0^+}$, even arrangements of two point interactions always converge to a single even point interaction. On the other hand, odd arrangements will converge to a single odd point interaction in that limit if, and only if, the two point interactions are odd themselves (i.e., $B=A_0=0$) \emph{or} $W=0$ in (\ref{Bodd})-(\ref{Wodd}). The last condition ($W=0$) \emph{includes} the case of an odd arrangement of two even point interactions ($W=A_1=0$).\footnote{In this case the single point limit degenerate to the free case, or the $\mathbb{\Lambda}$ matrix is just minus the identity matrix.} On the other hand, odd arrangements will converge to a {nontrivial} \emph{even} single point interaction in the ${\ell\to 0^+}$ limit  if, and only if, {$a=-d$} and {$\varphi\in\{0,\pi/2\}$} in (\ref{Lto0O}); this is an interesting and unexpected result, and illustrates the non trivial features of point interactions. In all other cases, the $\ell \to 0^+$ limit of odd arrangements of two point interactions does not have a well-defined parity.


\section{Critical, Supercritical, Bound, and Resonant, and Scattering States}

In this section, we will consider the existence of critical and supercritical ($E=+m$ and $E=-m$, respectively), bound states ( $-m< E<m$), and resonances for the Dirac equation (\ref{dir})-(\ref{d}). Critical (supercritical) states emerge when a bound state is absorbed (emitted) from (to) the continuum spectrum of the allowed energies, by varying the parameters of the interaction. Critical and supercritical states can be bound or quasi-bound states, depending on whether they are normalizable or not \cite{KNN22}. We will investigate resonances as the complex poles of the scattering $S$-matrix, which coincide with the complex energy solutions for purely outgoing scattering states.  We will start the section by classifying the states, and then we will study those very same states for symmetric arrangements.

\subsection{Classification of states}

\subsubsection{Critical states}

\emph{Critical states} are solutions of the  Dirac equation (\ref{dir})-(\ref{d}) with $E=+m$, \textit{i.e.}, they are solutions of 
\begin{eqnarray}\label{dircrit}
\left[i \gamma^1 \partial_1 -  m (\mathbb{1}  -\gamma^0 ) \right] \psi (x) = D[\psi](x) \; .
\end{eqnarray}

In the three regions separated by the barriers, the solutions (\ref{DirReg}) assume the form
\begin{equation}
\psi_j(x)=F_j \begin{pmatrix} 2 i m\,x \\1\end{pmatrix} + G_j \begin{pmatrix} 1 \\0\end{pmatrix}=\mathbb{P}_{{\mathrm{{crit}}}} \begin{pmatrix} F_j\\G_j\end{pmatrix}\,, \quad {j\in\{1,2,3\}},
\end{equation}
where 
$$
\mathbb{P}_{{\mathrm{{crit}}}}(x)=\begin{pmatrix} 2im\,x&1\\1&0 \end{pmatrix}\, .
$$

The requirement of boundedness of the spinor components {as} $x\to \pm \infty$ implies {$F_1=F_3=0$}. The coefficients $G_1$ and $G_3$ are connected {with one another }by a relation similar to (\ref{transfer}), with the matrix $\mathbb{P}$ in (\ref{Matrix M}) replaced by $\mathbb{P}_{{\mathrm{{crit}}}}$ (and with $F_1=F_3=0$). Thus,
\begin{eqnarray}\nonumber
\begin{pmatrix} 0\\G_3\end{pmatrix}&=&\mathbb{M}^{{\mathrm{{crit}}}}\left(\mathbb{\Lambda}_1,\mathbb{\Lambda}_2,x_1,x_2\right)\begin{pmatrix}0\\G_1\end{pmatrix}\,,\\
\label{mcrit} \\
\mathbb{M}^{{\mathrm{{crit}}}}\left(\mathbb{\Lambda}_1,\mathbb{\Lambda}_2,x_1,x_2\right)&=&\left[\mathbb{P}^{-1}_{{\mathrm{{crit}}}}(x_2)\mathbb{\Lambda}_2 \mathbb{P}_{{\mathrm{{crit}}}}(x_2) \right]\,\left[\mathbb{P}^{-1}_{{\mathrm{{crit}}}}(x_1)\mathbb{\Lambda}_1 \mathbb{P}_{{\mathrm{{crit}}}}(x_1)\right]\,.\nonumber
\end{eqnarray}

The {equation above} can be written as
\begin{equation}
\begin{pmatrix}{M}^{{\mathrm{{crit}}}}_{12}&0\\{M}^{{\mathrm{{crit}}}}_{22}&-1\end{pmatrix}\begin{pmatrix}G_1\\G_3\end{pmatrix}=0\,,
\end{equation}
where $M^{{\mathrm{{crit}}}}_{ij}$ is the $ij$-element of the matrix $\mathbb{M}^{{\mathrm{{crit}}}}$. {So,} the condition for nontrivial solutions is 
\begin{equation}\label{critcond}
{M}^{{\mathrm{{crit}}}}_{12}=0\,.
\end{equation}

Thus, the system will admit a nontrivial critical state only if the equation above is satisfied; in this case, the critical state is a \emph{quasi-bound} state, since the spinor $\psi(x)$ is not square integrable (it is a non-zero constant at infinity).

\textcolor{black}{
\subsubsection{Supercritical States}
}

\emph{Supercritical states} are solutions of the Dirac equation (\ref{dir})-(\ref{d}) with energy $E=-m$. 

Following the same steps as in the last section, we find the condition for the existence of nontrivial supercritical states to be
\begin{equation}\label{supercond}
{M}^{{\mathrm{{super}}}}_{12}=0\,,
\end{equation}
where, $M^{{\mathrm{{super}}}}_{ij}$ is the $ij$-element of the matrix $\mathbb{M}^{{\mathrm{{super}}}}$:
\begin{align}
    \mathbb{M}^{{\mathrm{{super}}}}\left(\mathbb{\Lambda}_1,\mathbb{\Lambda}_2,x_1,x_2\right)=\left[\mathbb{P}^{-1}_{{\mathrm{{super}}}}(x_2)\mathbb{\Lambda}_2 \mathbb{P}_{{\mathrm{{super}}}}(x_2) \right]\,\left[\mathbb{P}^{-1}_{{\mathrm{{super}}}}(x_1)\mathbb{\Lambda}_1 \mathbb{P}_{{\mathrm{{super}}}}(x_1)\right]
\end{align}
and
\begin{align}
    \mathbb{P}_{{\mathrm{{super}}}}(x)=\begin{pmatrix} 1&0\\-2im\,x&1 \end{pmatrix}.
\end{align}

Again, supercritical states are \emph{quasi-bound} states.

\subsubsection{Bound States}

\comjt{The conditions for the existence of bound states with $|E|<m$ follow directly from the formalism developed in section \ref{solution}. In this case   $k= i\kappa$ in (\ref{psi}), with $\kappa \equiv \sqrt{m^2 - E^2}$. The square-integrability of the Dirac spinor requires that $F_1=G_3=0$, since the terms in \eqref{psi} with these coefficients diverge at $|x|\to\infty$.  Using} the {boundary conditions} (\ref{bc}) {in} \textcolor{black}{ (\ref{transfer})}{, leads to} 
\begin{equation} \label{transferBound}
\left(
\begin{array}{c}
  F_3  \\
  0   \
\end{array}
\right)
= \mathbb{M}(\mathbb{\Lambda}_1,\mathbb{\Lambda}_2, i\kappa, x_1,x_2) 
\left(
\begin{array}{c}
  0  \\
  G_1  
\end{array}
\right)
\end{equation}
where the matrix $\mathbb{M}$ is given by (\ref{Matrix M}). {Nontrivial solutions of (\ref{transferBound}) exist \comjt{if, and }only if
\begin{equation}\label{boundcond}
\left. M_{22}\right|_{k\to i\kappa } =0\,.
\end{equation}

\textcolor{black}{
\subsubsection{Resonances}
}

\textcolor{black}{By imposing boundary conditions for a purely outgoing scattering state on the solutions (\ref{DirReg}) {we obtain $F_1=G_3=0$ and, thus,} exactly the same condition (\ref{boundcond}), {{but }with {$k$ changed to} $k=\sqrt{(E_R \pm i\,\Gamma/2)^2-m^2}$}}{, where $\Gamma>0$}. The real part $E_R$ of the complex energy solutions are the resonant energies. \footnote{The inverse of the imaginary part $\Gamma/2$ gives the mean life of the resonance \cite{GYa04}.}


\subsubsection{Scattering}
{For scattering} solutions with the particle incident from the left (no incident wave from the right), we can set $F_1 = 1$, $G_1=r(k)$, $F_3=t(k)$, and $G_3 = 0$, obtaining
\begin{eqnarray}
&&r(k) = -\frac{M_{21}}{M_{22}} \; , \label{r k}\\
&&t(k)= \frac{\det (\mathbb{M})}{M_{22}}\; , \label{t k}
\end{eqnarray} 
where $M_{jk}$ indicates the element $jk$ of the transfer matrix $\mathbb{M}$ {\eqref{Matrix M}}; $r(k)$ and $t(k)$ are the amplitudes of reflection and transmission, respectively. 

It is also interesting to write the form of the scattering matrix. If we denote the entries of the transfer matrix $\mathbb{M}(\mathbb{\Lambda}_1,\mathbb{\Lambda}_2, k, x_1,x_2) $ by $\{M_{ij}\}$, the form of the scattering $S$-matrix in terms of these entries is given by

\begin{equation}
    S = \frac{1}{M_{22}} \left( \begin{array}{cc} -M_{21} & 1 \\[2ex] \det \mathbb{M} & M_{12}   \end{array} \right)\,.
\end{equation}

Observe that the poles of the scattering $S$-matrix, which are given by  the same condition obtained by the purely outgoing conditions (namely $M_{22}=0$), coincide with those of the transmission coefficient \eqref{t k}. Therefore, we use this criterion to construct the graphics giving resonances in the cases to be considered later.


\subsection{Symmetric arrangements}

Now that we have defined all types of states to be considered, let us investigate the dependence of such states on the parameters of the interaction for some specific arrangements of well-defined symmetry.

\subsubsection{Even arrangements}

An even arrangement of the two barriers is characterized by $\mathbb{\Lambda}$-matrices as given by (\ref{Leven}). Substituting these matrices into (\ref{critcond}), (\ref{supercond}), and (\ref{boundcond}), we obtain the conditions below for critical,  supercritical, and bound states.

\begin{itemize}

\item[\emph{i)} ] \textcolor{black}{\emph{Critical states}: Condition (\ref{critcond}) simplifies to $c( a +  c \ell m)=0$. This implies 
\begin{equation}\label{criteven}
c=0\qquad\text{or}\qquad {a+ c\, m\ell = 0}\,.
\end{equation}
}
\item[\emph{ii)} ] \textcolor{black}{\emph{Supercritical states}: Condition (\ref{supercond}) gives $ b( d +  b \ell m)=0$, which, in turn, implies 
\begin{equation}\label{supereven}
b=0\qquad\text{or}\qquad {d + b\, m\ell=0}\,.
\end{equation}
}

\item[\emph{iii)}] \emph{Bound states and resonances}: Condition (\ref{boundcond}) now reads   
\begin{equation} \label{EnEven}
{\left( a+d+\frac{\kappa}{m+E} b +\frac{m+E}{\kappa}c\right)} e^{\kappa \ell} = \pm {\left( a-d+\frac{\kappa}{m+E} b -\frac{m+E}{\kappa}c\right)}
\end{equation}
Real energy solutions for this equation give the bound-state energies, and the real part of the complex energy solutions give the resonant energies. 

\end{itemize}

We observe that the phase $\varphi$ in (\ref{Leven}) does not have any effect on the spectrum of states that we are considering.

\subsubsection{Odd Arrangements}

\textcolor{black}{An odd arrangement of the two barriers is characterized by $\mathbb{\Lambda}$-matrices as given by (\ref{Lodd}). By using (\ref{critcond}), (\ref{supercond}){,} and (\ref{boundcond}) with these matrices  we obtain the conditions below for critical,  supercritical{,} and bound states.}

\begin{itemize}

\item[\emph{i)} ] \textcolor{black}{\emph{Critical states}: Condition (\ref{critcond}) gives $\,c (d-a -2 c  \ell m)=0$, which implies that 
\begin{equation}\label{critodd}
c=0\qquad\text{or}\qquad {(d-a)- 2m\ell\, c=0}\,.
\end{equation}
}
\item[\emph{ii)} ] \textcolor{black}{\emph{Supercritical states}: Condition (\ref{supercond}) gives $ b(a-d-2bm\ell )=0$ which gives  
\begin{equation}\label{superodd}
b=0\qquad\text{or}\qquad {(d-a)+ 2m\ell\, b=0}\,.
\end{equation}
}

\item[\emph{iii)}] \textcolor{black}{\emph{Bound states and resonances}: Condition (\ref{boundcond}) now reads
\begin{equation} 
{\left( a-d+\frac{\kappa}{m+E} b -\frac{m+E}{\kappa}c\right)} e^{-\kappa \ell} = \pm \sqrt{{\left( \frac{\kappa}{m+E} b +\frac{m+E}{\kappa}c\right)}^2 -(a+d)^2}\, .\label{oddCond1}
\end{equation}
}
For bound states the left{-}hand side (lhs) must be real, and this condition will have solutions if and only if
\begin{equation} \label{OddCond2}
\left| \frac{\kappa}{m+E} b +\frac{m+E}{\kappa}c\right|\geq |a+d|\,.
\end{equation}
We observe that single odd interactions necessarily have $b=c=0$ (or, in terms of the physical strengths, $B=A_0=0$ - see Appendix \ref{app strengths} or \cite{BLM24}). It is immediately clear from (\ref{OddCond2}) that an odd arrangement of two single \emph{odd} point interactions does not possess bound states.

Resonant energies will be given by the real part of the complex energy solutions of (\ref{oddCond1}).

\end{itemize}

Finally, as in even arrangements, the phase $\varphi$ in (\ref{Lodd}) is irrelevant to calculate the critical, supercritical, bound, and resonant energies for odd arrangements. 

In the next section, we will consider some particular cases of even and odd arrangements of two point interactions.

%
%

\section{Some Particular Cases}

In this section we will consider some specific even and odd arrangements of a pair of point interactions. The cases we will consider are: \emph{i})  equal mixtures of scalar and electrostatic strengths (which in the non-relativistic limit gives a pair of  ``$\delta$ interactions"); \emph{ii})  inverted mixtures of scalar and electrostatic interactions (resp. a pair of non-local``$\delta^\prime$ interactions"); \emph{iii}) a pair of pseudoscalar interactions (two ``local $\delta^ \prime$ interactions") and \emph{iv}) a pair of magnetic interactions.

\subsection{Even arrangements}
\label{evensubsec}

\subsubsection{Equal \comlm{Mixtures} of Scalar and Electrostatic Point Interactions}

We consider the even arrangement (\ref{Leven}) with each point interaction an equal mixture of electrostatic and scalar point interactions, \textit{i.e.}, \comcb{$A_0=B$ arbitrary} and $A_1=W=0$. It follows from Appendix \ref{app strengths} that, in terms of the $\Lambda$-parameters, this corresponds to $\varphi=b=0$, $a=d=1$ and $c=2A_0=2B=A_0+B$ -- in the non-relativistic limit this equals to two identical $\delta$ interactions (for a brief review of the non-relativistic limit for all interactions considered in this work, see Appendix \ref{app NR}). Let us now investigate whether this double barrier admits critical, supercritical, or bound states.

\begin{itemize}

\item \emph{Critical and supercritical states}. Substituting {the choice of parameters corresponding to an even two delta configuration} in (\ref{criteven}), we {find} critical states for 
\begin{equation}\label{crit2delta}
c=0\qquad\text{or}\qquad c=-\frac{1}{m \ell}\, .
\end{equation}
The first case ($c=0$) corresponds to the trivial case of a free particle at rest. On the other hand, since $b=0$, condition (\ref{supereven}) for supercritical states is satisfied for \emph{any} value of $c$ and there is no absorption or emission of a bound state at the supercritical energy $E=-m$.   

\item \emph{Bound states}. The conditions for bound states, (\ref{EnEven}), in this case are reduced to 
\begin{equation} 
{\left( 2 +\frac{m+E}{\kappa}c\right)} e^{\kappa \ell} = \mp \frac{m+E}{\kappa}c
\end{equation}
or, after a slight manipulation,
\begin{align} 
\sqrt{\frac{m - E}{m+E} } {=}  -\frac{c}{2}{\left(1\pm e^{-\kappa \ell} \right)=}-A_0 {\left(1\pm e^{-\kappa \ell} \right)}\, ,\label{BSEvenDelta}
\end{align}
where we {have made} the replacement $c=2A_0$ in the last line. It is clear that bound states can exist only for $A_0<0$. In the equations above, the solution with the plus sign will be denoted by $E_+$, and it stands for the ground state; accordingly, $E_-$ denotes the excited state. 
\begin{figure}[h!]
\includegraphics[scale=0.9]{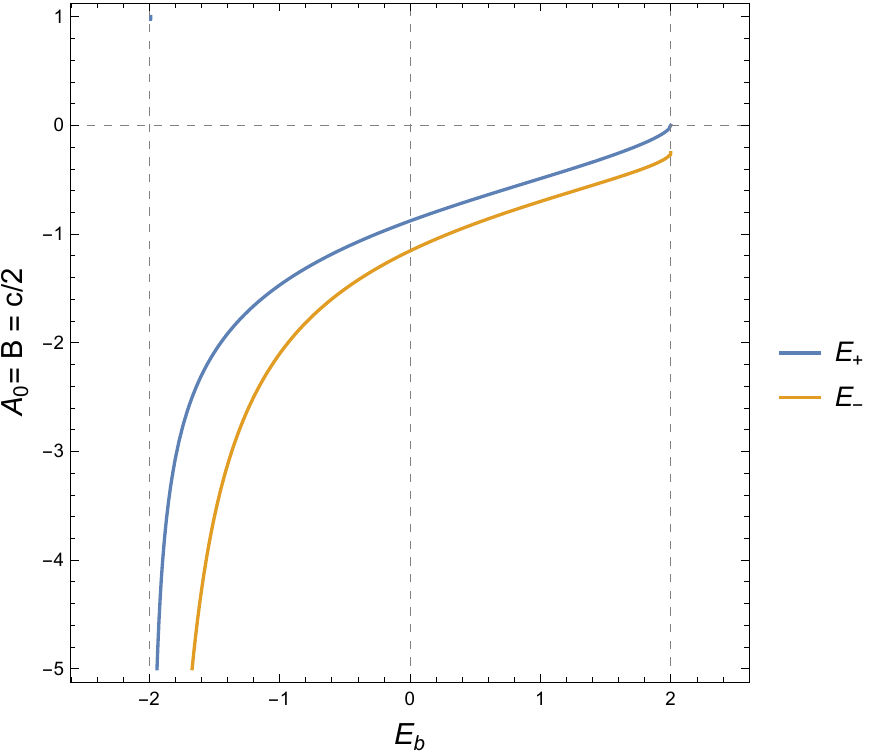}
\caption{ \label{figBS2deltasEven} Relationship between bound-state energies and the interaction strength constant $A_0=B=\frac{c}{2}$ for an even arrangement of two identical interactions, each one consisting of an equal mixture of an electrostatic and a scalar interaction. In the figure, $m=2$ and $\ell=1$. The blue curve corresponds to the ground state, and the yellow {one} to the excited state. There are two critical states ($E=+m$), corresponding to $A_0=0$ (free particle at rest) and $A_0=-\frac{1}{2m\ell}=-\frac{1}{4}$. When $A_0$ goes from $0$ to $-\infty$ a bound state (the ground state $E_+$) is absorbed from the continuum at $A_0=0$, and another one (the excited state $E_-$) is absorbed from the continuum when $A_0=-\frac{1}{2m\ell}$. For $A_0<-\frac{1}{2m\ell}$ the system maintains these two bound states{, with}  $E_\pm\to -m$ as $A_0 \to-\infty$.}
\end{figure}

Figure \ref{figBS2deltasEven} shows a graph for the bound states. When $A_0=B=\frac{c}{2}$ varies from positive to negative values, a bound state (the ground state $E_+$) is absorbed from the continuum as $A_0$ crosses {the boundary} $A_0=0$. Furthermore, this bound state is maintained (with decreasing energy) for every negative value of $A_0$. When $A_0$ varies from $A_0=0$ towards $-\infty$, another bound state (the excited state $E_-$) is captured at the critical energy $E=+m$ when $A_0=-\frac{1}{2m\ell}$. This bound state is also maintained as $A_0\to-\infty$. By varying the parameter $A_0$, there is no absorption (emission) of a bound state from (to) the continuum at the supercritical energy ($E=-m$); as we have seen above, the supercritical state is admitted for every value of $A_0$. 

It follows from (\ref{BSEvenDelta}) that excited states exist only for a separation between the point interaction larger than the critical $\ell_{\rm cr} = 1/(m|c|)$. In the limit $\ell\to\infty$ the energies become degenerate and tend to the bound state energy of a single interaction, $E_{\pm}\overset{\ell\to \infty}{\longrightarrow} \frac{(1-c^2/4)}{(1+c^2/4)}m$ -- in Figure \ref{figBS2deltasEven} both curves would coalesce into a single one, with both the ground and the excited states captured as critical states at $A_0=0$. On the other hand, in the limit $l\to 0^+$ the capture of the excited state occurs at $A_0\to -\infty$ and there is one bound state for a single point interaction with twice the strength (\textit{i.e.}, $E_+\overset{\ell\to 0^+}{\longrightarrow} \frac{(1-c^2)}{(1+c^2)}m $) at $x=0$, in agreement with (\ref{Lto0E}).

\item \emph{Resonances}. To find the complex energy solutions for (\ref{BSEvenDelta}) it is convenient to write it in the form (here we replace $\kappa= -i k$, for purely outgoing scattering solutions) 
%
\begin{equation} \label{ResEvenDelta}
\frac{i k}{(m+E)\left(1\pm e^{i k\ell}\right)}=  \frac{c}{2}=A_0=B.
\end{equation}
Now the lhs of this equation may be complex, since $E=E_R \pm i \Gamma/2$ \comlm{($\Gamma >0$)} and $k=\sqrt{\left(E_R\pm i \Gamma/2\right)^2-m^2}$. The imaginary part of the lhs of this complex equation must be zero, {because} $c$ (that is, $A_0$ and $B$) is real. Moreover, the imaginary part of this equation does not depend on the strength $A_0$ (equivalently, $B$ or $\frac{c}{2}$). Figure \ref{figRES2deltasEven} shows a plot of the complex energies that solve (\ref{ResEvenDelta}) \footnote{As observed in the introduction, the energies characterizing the resonances always come in complex pairs. However, in Figure \ref{figRES2deltasEven}, and all subsequent resonance figures, we only observe half of these pairs due to the fact that we find the poles of the $S$-matrix only for the principal component root, \textit{i.e.}, for $M_{22}(+k)=0$. The other half of the resonances is, therefore, trivially obtained by taking the complex conjugate energies.}. We observe that in the limit $A_0 \to \pm \infty$, the double barrier becomes impermeable, and the complex energies that solve the equation become all real; these energies correspond to the infinite discrete set of energies allowed for the (or antiparticle) confined in the region $[-\ell,\ell]$. As the interaction strength $A_0$ increases in absolute value, the complex energies move alongside the ``U" shaped blue and orange curves towards this discrete set of real energies. For $A_0\leq 0$ and increasing in absolute value, the system captures a first bound state at $A_0=0$, and at $A_0=-\frac{1}{2m\ell}$ ($A_0=-0.25$ in the plot) the system captures the second bound state. As $A_0\to -\infty$, both bound states approach the energy $-m$. For any value of the strength $A_0$, there is a supercritical state at $E=-m$, which means that no bound state is emitted or absorbed at $E=-m$.

\begin{figure}[h!]
\includegraphics[scale=1.3]{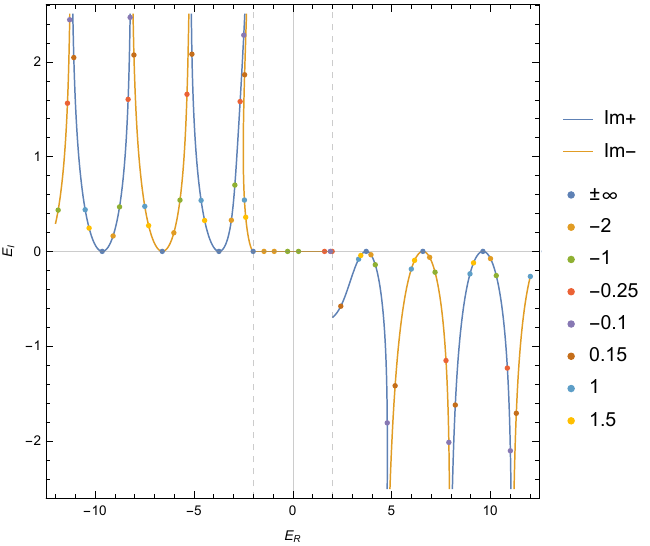}
\caption{ \label{figRES2deltasEven} Complex energies corresponding to purely outgoing solutions of the Dirac equation with an even arrangement of two identical point interactions, each one being an equal mixture of an electrostatic and a scalar point interactions. In the figure, $m=2$ and $\ell=1$. The colored points are the complex energies that solve both the real and imaginary parts of the complex equation (\ref{ResEvenDelta}) for the values of $A_0$ (equivalently, $B$) indicated in the legend. The blue and orange curves correspond, respectively, to solutions of the imaginary part of (\ref{ResEvenDelta}) with the plus (Im+) and minus (Im-) sign.   For any value of $A_0$ there is a supercritical state at $E=-m$. The colored points on the ``U-shaped" curves move towards the real axis as $A_0$ increases in absolute value, approaching a set of values of energy corresponding to a particle (or antiparticle) confined in the inner region of the double barrier interaction. The bound state energies in the region $-m<E_R<m$ move to the left as $A_0$ increases in absolute value towards negative values.}
\end{figure}

\end{itemize}

\subsubsection{Inverted Mixtures of Scalar and Electrostatic Interactions}

In this case we choose $A_0=-B$, $A_1=W=0$, which corresponds to $a=d=1, \varphi=c=0$ and $b=2A_0=-2B=A_0-B$  in (\ref{Leven}). In the non relativistic limit each of the two point  interactions corresponds to a non-local $\delta^\prime$ interaction. 

\begin{itemize}

\item \emph{Critical and supercritical states}. Since in this case $c=0$, equation (\ref{criteven}) is trivially satisfied for any value of the parameter $b$. Therefore, the system admits a critical state for every value of $b$ and no bound state is emitted or absorbed at critical energy, $E=+m$, by varying $b$. On the other hand, condition (\ref{supereven}) for supercritical states gives 
\begin{equation}\label{super2deltaP}
b=0\qquad\text{or}\qquad b=-\frac{1}{m\ell}\, .
\end{equation}
The case $b=0$ corresponds to the trivial case of a free (anti-)particle at rest.

\item \emph{Bound states}. In this case, after some manipulation, equation (\ref{EnEven}) results
\begin{align}
\sqrt{\frac{m+E}{m-E} } =  -\frac{b}{2}{\left(1\mp e^{-\kappa \ell} \right)}= -A_0{\left(1\mp e^{-\kappa \ell} \right)}\, ,\label{BSEvenNLDeltaP}
\end{align}
where in the last step we have replaced $b=2A_0=-2B$. Note that this equation may be obtained from (\ref{BSEvenDelta}) by making the substitution $E\to -E$; in fact, the same substitution explains the critical and supercritical states. Therefore, all the results for bound states and resonances for the present case follow from the results for an even arrangement of two point interactions, each an equal mixture of electrostatic and scalar interactions just by making $E\to -E$, and the even arrangement of two inverted mixtures of scalar and electrostatic interactions behaves as an even arrangement of two equal mixtures of the same interactions acting on the ``antiparticle sector''.

\end{itemize}

%
%

\subsubsection{Two Pseudoscalar Interactions}

The even arrangement of pseudoscalar interactions is obtained with the choice
\begin{equation}
A_0=A_1=B= 0 {\qquad}  {\rm and } \qquad W\comcb{\,\,\,\mbox{arbitrary}} ;
\end{equation}
in (\ref{Beven})-(\ref{Weven}), which, due to (\ref{ap})-(\ref{dp}), we have: 
\begin{equation} \label{Pseudo}
d=\frac{1}{a}=\frac{2+W}{2-W}{\qquad}  {\rm and} \qquad \varphi=b=c=0\, .
\end{equation}
In the non relativistic limit, each of the two point interactions corresponds to the so-called ``local" $\delta'$ interaction. In the relationships above, both $W=\pm 2$ cause  $a$ or $d$ to diverge. These cases correspond to a pair of \emph{impermeable} point interactions (see eq. (55) in \cite{BLM24}). The boundary conditions at the points $x=-\frac{\ell}{2}$ and $x=+\frac{\ell}{2}$, in this case,  depend on the value of $W$ being $+2$ or $-2$. If $W=-2$, we have $(1\quad 0)\,\psi\left(\pm\frac{\ell}{2}^-\right)=0$ and $(0\quad1)\,\psi\left(\pm\frac{\ell}{2}^+\right)=0$. If, on the other hand, $W=+2$ we have $(0\quad1)\,\psi\left(\pm\frac{\ell}{2}^-\right)=0$ and $(1\quad0)\,\psi\left(\pm\frac{\ell}{2}^+\right)=0$. 

\begin{itemize}

\item \emph{Critical and supercritical states}. Conditions (\ref{criteven}) and (\ref{supereven}) for the existence of critical and supercritical states, respectively, are always satisfied for any value of $a=\frac{1}{d}$, since $b=c=0$. This means that, by varying $a=\frac{1}{d}$  (or, equivalently, by varying $W$), there is no absorption (emission) from (to) the continuum to (from) bound states.

\item{\emph{Bound states}. Equation (\ref{EnEven}) can now be written as
\begin{eqnarray} \label{PseudoBSoriginal} 
e^{\kappa \ell} &=& \pm\frac{a^2-1}{a^2+1}=\pm\frac{4 W}{4+W^2}\\ \label{PseudoBS}
&=& \frac{4 |W|}{4+W^2}\,,
\end{eqnarray} 
where in the second line we have taken into account that the lhs is strictly positive, and thus to find bound states we \emph{must} choose the plus sign when $W> 0$ and the minus sign when $W<0$. On the other hand, since $\kappa >0$, the lhs of the {equation above} is strictly greater than unity, whereas $0\leq \frac{4 |W|}{4+W^2} \leq 1$. Therefore, the {equation above} has no solution, and no bound states exist. This is consistent with the fact that a single pseudoscalar point interaction does not admit bound states, since in the limits $\ell\to \infty$ or ${\ell\to 0^+}$ the even arrangement would have the same structure for bound states as a single point interaction.}

\item{\emph{Resonances}. For complex energies, equation (\ref{PseudoBSoriginal}), with $\kappa\to -i k$, may have solutions for both signs, and can be written more conveniently as 
\begin{equation}\label{resPS}
e^{-2 i k \ell}= \left(\frac{a-d}{a+d}\right)^2=\left(\frac{4 W}{4+W^2}\right)^2\,. 
\end{equation}
This equation has the symmetries $W\to -W$ and $W\to \frac{4}{W}$. So, it is sufficient to investigate its complex energy solutions for $W\in [0,2]$.
Figure \ref{figRES2localdeltasEven} shows the {behavior} of the complex energy solutions of (\ref{resPS}). The blue curve corresponds to the solution of the imaginary part of the equation, which does not depend on the strength $W$. The colored points are the complex energies that solve both the real and complex parts of the complex equation for {some particular} values of {$W\in[0,2]$}, shown in the legend. As {$|W|$} increases within this interval{,} the complex energies {move towards the real axis alongside the blue curves}. When $|W|=2${,} the double barrier becomes impermeable {and} the set of real solutions corresponds to the allowed energies for a particle (or antiparticle) confined within the impenetrable walls at  $-\ell$ and $\ell$.}
\begin{figure}[h!]
\includegraphics[scale=1.3]{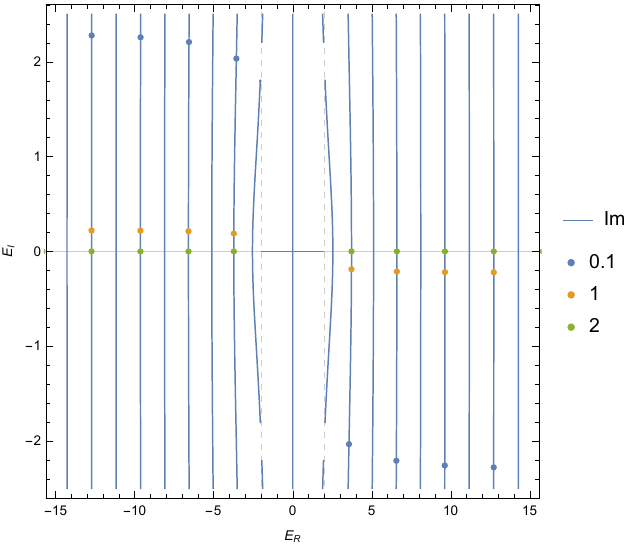}
\caption{ \label{figRES2localdeltasEven} Complex energies corresponding to purely outgoing solutions of the Dirac equation with an even arrangement of two pseudoscalar interactions. In the figure, $m=2$ and $\ell=1$. The blue curves correspond to solutions for the imaginary part of the {complex equation (\ref{resPS}). The colored points are the complex energies that solve the complex equation with the values of $W$ indicated on the right. The real energies for values $W=\pm 2$ correspond to the set of energies for a particle (or antiparticle) confined within the two impenetrable barriers. When $W$ decreases in absolute value from $|W|=2$ the complex energies move away from the real axis. The same complex energies solve the equation (\ref{resPS}) with $W\to \frac{4}{W}$}.}
\end{figure}

\end{itemize}

%
%
\subsubsection{Two magnetostatic interactions (two singular gauge fields)}

In this case, we should consider the following parameters in (\ref{Beven})-(\ref{Weven}):
\begin{equation}
A_0=B= W=0\,  {\qquad}  {\rm and } \qquad A_1 \comcb{\,\,\,\mbox{arbitrary}} ;
\end{equation}

which, due to (\ref{ap})-(\ref{dp}), imply 
\textcolor{black}{
\begin{equation} \label{PureMag}
a=d=-\text{sign}\, {(A_1)} \, , \qquad b=c=0 \qquad  {\rm and } \qquad \varphi=\tan^{-1} {\left(\frac{4 A_1}{A_1^2-4}\right)}\,\quad \varphi\in [0,\pi)\,.
\end{equation}
}
That is, the only non-vanishing parameter in (\ref{Leven}) is the phase parameter $\varphi$. As mentioned in the last section, this phase has no effect on the existence of critical or supercritical states, as well over the energies of bound states and resonances. Therefore, in what concerns these features, in this case the system behaves as a free particle.

%
%
\subsubsection{Two Scalar Interactions}

Here, we choose \comcb{$B$ arbitrary}, $A_0=A_1=W=0$ in (\ref{Beven})-(\ref{Weven}). Thus, from (\ref{ap})-(\ref{dp}), we have
$$
d=a=\frac{4+B^2}{4-B^2},\qquad b=c=\frac{4B}{4-B^2}\, .
$$

For $B=\pm 2$, each one of the point barriers is \emph{impermeable}. 

\begin{itemize}

\item[]\emph{Critical and supercritical states}. Equations (\ref{criteven}) and (\ref{supereven}) give the same conditions for both critical and supercritical states, which are 
\begin{equation}\label{PScritsuper0}
B=0\qquad\text{or}\qquad \frac{4 B}{4+B^2}=-\frac{1}{m\ell}\,.
\end{equation}

The first of the conditions above corresponds to the free case while the second one gives 
\begin{equation}\label{PScritsuper1}
B_\pm=-2\left(m\ell\pm\sqrt{(m\ell)^2-1}\right)\, ,
\end{equation}
which have real solutions if, and only if, $\ell\geq \frac{1}{m}$. The two solutions in (\ref{PScritsuper1}) are related by the transformation $B\to\frac{4}{B}$, which is a symmetry of the second of equations (\ref{PScritsuper0}). Thus, each value of $B$ satisfying the {conditions above} allows for both a critical and a supercritical state. As we will see below, at all values of $B$ satisfying (\ref{PScritsuper0}) there is simultaneous absorption and emission of a bound state at $E=-m$ and $E=+m$.

\item[]\emph{Bound states}. Conditions (\ref{EnEven}) for this case can be written as
\begin{equation}\label{EnEvenPS}
\frac{a}{b}=\frac{4+B^2}{4B}=\frac{-m\pm \left|E\right| \,e^{-\kappa \ell}}{\kappa}\, ,
\end{equation}
which also has the symmetry $B\to \frac{4}{B}$, and $E\to -E$. Thus, to investigate the structure of the bound states and resonances it is enough to consider $B\in [-2,2]$.

\textcolor{black}{Figure \ref{figBS2ScalarEven} shows the relationship between $B$ and the bound states. In this Figure the blue and the orange curves (indicated by $E_+$ and $E_-$) correspond respectively to the plus and minus signs in (\ref{EnEvenPS}). For any negative value of $B$ {{the} orange curve gives two bound states. When $B$ is in between the roots $B_\pm$ of (\ref{PScritsuper1}) there are two additional bound states, given by the blue curves.} There are two critical and two supercritical states corresponding to the values of $B_\pm$ satisfying (\ref{PScritsuper1}). There are also one supercritical and one critical state when $B=0$, which correspond to a free antiparticle and a free particle at rest, respectively. When $B$ crosses $0$ from above, two bound states (orange curve) are absorbed from the continuum {at} $E=\pm m$. By decreasing {$B$} from zero towards negative values, two new bound states are absorbed from the continuum at $E=\pm m$ when $B$ takes the value $B_-$ (blue curve) -- these are later emitted into the continuum when $B$ takes the value $B_+=\frac{4}{B_-}$. For $B=-2$ the two point barriers become impermeable and the system admits two nonzero energies in the interval $[-m,m]$ (blue curve) for a particle confined in between the two barriers (in which case the spinor vanishes identically outside $\left[-\frac{\ell}{2},\frac{\ell}{2}\right]$) as well as a state with energy $E=0$ (orange curve), with the spinor vanishing between the barriers and being nonzero outside them.\footnote{The state with $E=0$ in this case is analogous to the single bound state admissible for a single, scalar, impermeable barrier, since the double impermeable barrier ``cuts{ off}" the region $\left(-\frac{\ell}{2}, \frac{\ell}{2}\right)$ from the real line.} 
If  $\ell< \frac{l}{m}$, only the two bound states given by the orange curve exist, for any value of \comlm{$B<0$}. In the limit $\ell \to 0^+${,} the orange curve yields the bound states of a single, scalar, point interaction with strength $\tilde{B}=\frac{8B}{B^4+4}$. Finally, for $\ell\to \infty${,} the two point interactions become independent and the orange and blue curves coalesce to a single curve {giving} the bound states of a single scalar point interaction with strength $B$.}

\begin{figure}[h!]
\includegraphics[scale=0.9]{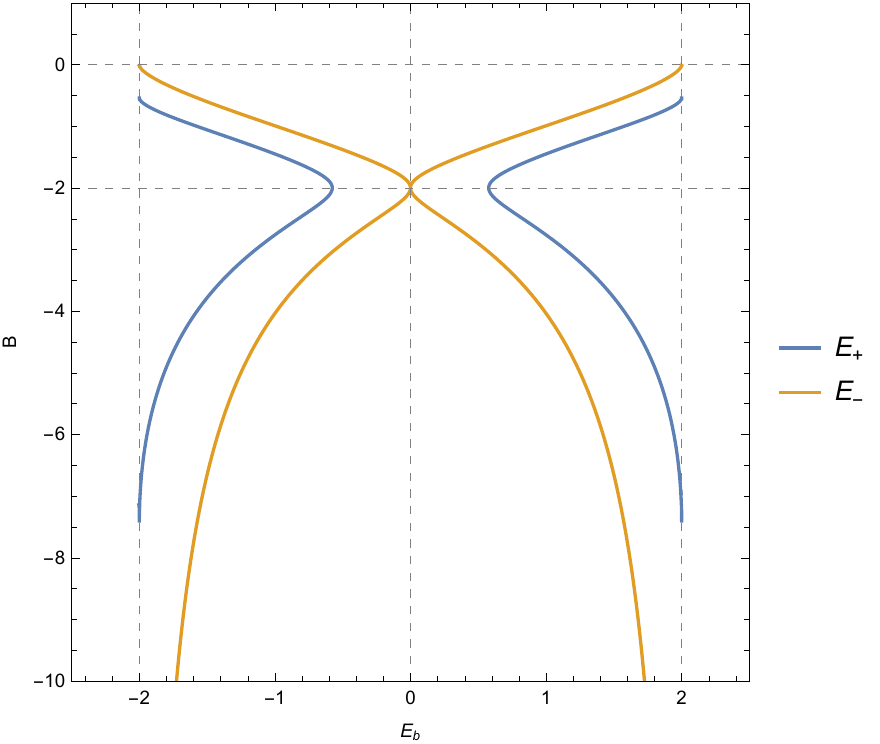}
\caption{ \label{figBS2ScalarEven} Relationship between bound-state energies and the strength constant $B$ for an even arrangement of two identical scalar interactions. In the figure, $m=2$ and $\ell=1$. The orange curve corresponds to the minus sign  in equation (\ref{EnEvenPS}), while the blue one corresponds to the plus sign .}
\end{figure}

\item[]\emph{Resonances}. Equation (\ref{EnEvenPS}), with $\kappa=-{i} k$, can be suitably rewritten as
\begin{equation}\label{resScalar}
    \frac{4+B^2}{4B}=\frac{-m\pm E \,e^{i k \ell}}{-i k}\, .
\end{equation}
Figure \ref{figRES2PureScalarEven} shows the complex energy solutions for this equation. The solid curves correspond to {solutions} of the imaginary part of that equation, which does not depend on the interaction strength $B$. The colored points correspond to the solutions for the whole complex equation for several values of the strength $B\in [-2,2]$. For $B=\pm 2$, the two barriers become impenetrable and the complex energies become real, with the set of discrete real energies for the cases $B=-2$ and $B=+2$ differing because the boundary conditions at $-\frac{\ell}{2}$ and $+\frac{\ell}{2}$ are different in the two cases. This set of real energies correspond to the allowed energies for a particle (or antiparticle) confined in between the two barriers. With $B$  within the interval $[-2,2]$ but decreasing in absolute value, the complex energies move along the ``U" shaped curves away from the real axis. There {is} the absorption/emission of two bound states when $B=0$ and another two when  $B=B_-$ ($B_-\approx -.54$ in the plot). The same complex solutions solve equation (\ref{resScalar}) with $B\to \frac{4}{B}$.
\begin{figure}[h!]
\includegraphics[scale=1.3]{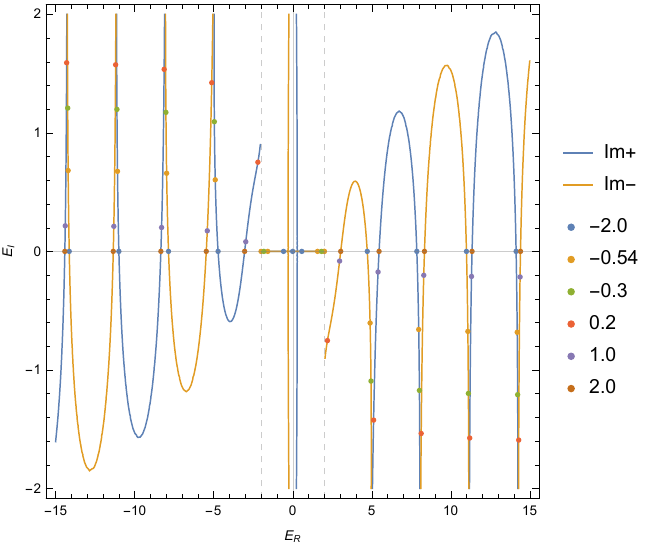}
\caption{ \label{figRES2PureScalarEven} Complex energies corresponding to purely outgoing solutions of the Dirac equation with an even arrangement of two scalar interactions. In the figure, $m=2$ and $\ell=1$. The blue and the orange curves correspond to complex energy solutions for the imaginary part of the complex equation (\ref{EnEvenPS}), with the plus and minus sign, respectively. The colored points are the complex energies that solve the complex equation for several values of the strength $B\in [-2,2]$. For $B=\pm 2${,} the two barriers are impenetrable and the real solutions for the energies  correspond to a particle (or antiparticle) confined in between the barriers (with different boundary conditions depending on the sign of $B$). The set of real energies in the interval $[-m,m]$ corresponds to the solutions of the equation for bound states (\ref{EnEvenPS}).}
\end{figure}

\end{itemize}

%
%
\subsubsection{Two Electrostatic Interactions}

As the last special case of even arrangements, we will consider two electrostatic point interactions, which are obtained by assuming \comcb{$A_0$ arbitrary} and $B=A_1=W =0$ in (\ref{Beven})-(\ref{Weven}). Using (\ref{ap})-(\ref{dp}), this corresponds to take
$$
\varphi=0,\qquad a=d=\frac{4-A_0^2}{4+A_0^2},\qquad c=-b=\frac{4A_0}{4+A_0^2} 
$$
in (\ref{Leven}). For any finite or infinite value of the strength $A_0$, each electrostatic point interaction is always permeable, see (\ref{permcond}). This has been interpreted as a signature of the Klein effect for point interactions, since it implies that it is impossible to confine a particle (or antiparticle) in one side of a barrier by using only electrostatic interactions \cite{MSt87}.

\begin{itemize}

\item \emph{Critical and supercritical states}. Equation (\ref{criteven}) for the existence of critical states now reads, in terms of the strength $A_0$, as\footnote{\textcolor{black}{If $A_0=\pm 2${,} we must have $a=0$ and the condition (\ref{criteven}) would imply $c=0$, which is impossible to satisfy given that $ad-bc=1$.}}
\begin{equation}\label{PEcritcondEven0}
A_0=0\qquad\text{or}\qquad \frac{4A_0}{4-A_0^2}=-\frac{1}{m\ell},\quad A_0\neq \pm 2.
\end{equation}
The first of these equations corresponds to the free case, and the second one has solutions
\begin{equation}\label{PEcritcondEven1}
A^{\mathrm{crit}}_{0\pm}=2\left(m\ell \pm  \sqrt{(m\ell)^2+1}\right).
\end{equation}
The solutions above satisfy $A^{\mathrm{{crit}}}_{0+}=-\frac{4}{A^{\mathrm{{crit}}}_{0-}}$, which is a symmetry of the second condition in (\ref{PEcritcondEven0}). Thus, for any value of the strength $A_0$ which admits a critical state, there will be a corresponding strength  with inverted sign, $-\frac{4}{A_0}$, which also admits a critical state.  
\\Similarly, equation (\ref{supereven}) for the existence of supercritical states gives
$$
A_0=0\qquad \text{or}\qquad A^{\mathrm{{super}}}_{0\pm}=2\left(-m\ell \pm  \sqrt{(m\ell)^2+1}\right)\; .
$$
Here too, we have the symmetry $A^{\mathrm{{super}}}_{0+}\to -4/A^{\mathrm{{super}}}_{0-}$.

\item \emph{Bound states}. Equation (\ref{EnEven}) now can be written as
\begin{equation}\label{EnEvenPE}
\frac{a}{b}=\frac{A_0^2-4}{4 A_0}=\frac{E\pm m\, e^{-\kappa\ell}}{\kappa}\, ,
\end{equation}
which, as expected, is also invariant under the transformation $A_0\to -\frac{4}{A_0}$. Figure \ref{figBS2ElectrostaticEven} shows the relationship between bound-state energies and the strength constant $A_0$ for an even arrangement of two identical electrostatic interactions, with the parameters $m=2$ and $\ell=1$. The blue curve, denoted by $E_+$ and the orange one, denoted by $E_-$, are the solutions corresponding, respectively to the plus and minus signs in (\ref{EnEvenPE}). We observe that there is at least one bound state for any value of $A_0$, and it follows from the symmetry $A_0\to -\frac{4}{A_0}$ that for any bound state associated to the value $A_0$ there corresponds a bound state, of same energy, associated to $-\frac{4}{A_0}$ -- as a consequence, the blue curves in the figure can be obtained from the orange ones through the transformation $E\to -E, A_0\to -A_0$, and vice-versa. 
\\The critical (supercritical) states are admitted when the strength $A_0$ takes the values {$0$,}  $A^{{\mathrm{crit}}({\mathrm{super}})}_{0+}$ and $A^{{\mathrm{crit}}({\mathrm{super}})}_{0-}=-\frac{4}{A^{{\mathrm{crit}}({\mathrm{super}})}_{0+}}$.  When the strength $A_0$ crosses $A_0=0$ from below, a bound state is absorbed from the continuum at the energy $E=-m$ and, continuing to increase $A_0$, a new bound state is absorbed from the continuum when $A_0=A^{{\mathrm{{super}}}}_{0+}$ at $E=-m$, becoming the new ground state. As $A_0$ continues to increase, the excited state (orange curve) is emitted into the continuum at $E=+m$ when $A_0=A^{{\mathrm{{crit}}}}_{0+}$. When $A_0>A^{{\mathrm{{crit}}}}_{0+}$ there is always a single bound state. A similar behavior occurs when $A_0<0$ becomes more negative: a bound state is absorbed at $E=+m$ when $A_0=0$, another bound state occurs at $E=+m$ when $A_0=A^{{\mathrm{{crit}}}}_{0-}$, and a bound state is emitted into the continuum at $E=-m$ when $A_0=A^{{\mathrm{{super}}}}_{0-}$; when $A_0<A^{{\mathrm{{super}}}}_{0-}$ the system admits a single bound state. When $\ell\to \infty$ the blue and the orange curves coalesce, corresponding to a single electrostatic point interaction with strength $A_0$, as can be seen from (\ref{EnEvenPE}). On the other hand, when ${\ell\to 0^+}$ the structure of the bound states is the same as that for a single electrostatic point interaction with strength $\tilde{A}_0=\frac{8 A_0}{4-A_0^2}$, which always admit a single bound state (in this case the bound states are emitted/absorbed when $A_0$ crosses $A_0=0,\pm 2$, and the emission of a bound stated at $E=+m $ is always simultaneous to the absorption of another one at $E=-m$, and vice-versa).

\begin{figure}[h!]
\includegraphics[scale=0.9]{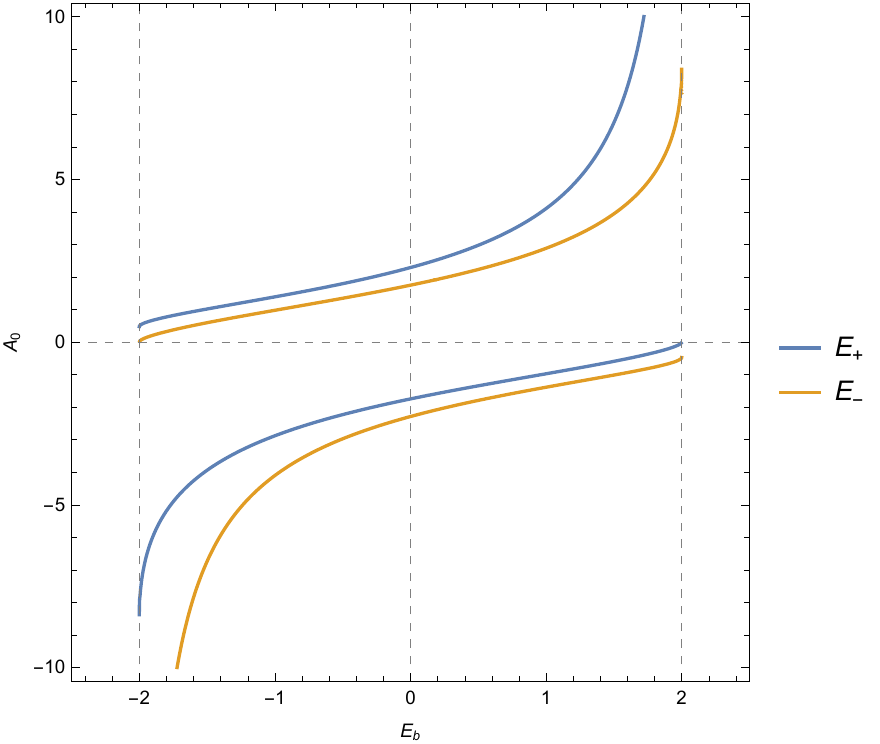}
\caption{ \label{figBS2ElectrostaticEven} Relationship between bound-state energies and the strength constant $A_0$ for an even arrangement of two identical electrostatic interactions. In the figure, $m=2$ and $\ell=1$. The orange curve corresponds to the minus sign in equation (\ref{EnEvenPE}), while the blue one corresponds to the plus sign.}
\end{figure}

\item \emph{Resonances}. Now we seek complex energy solutions of equation (\ref{EnEvenPE}), with $\kappa\to -{i}k$. From the symmetry $A_0\to -\frac{4}{A_0}$ it would be enough to consider $A_0 \in [0,+\infty)$ to investigate the structure of the resonances.  Figure \ref{figRES2PureElectrostaticEven} shows the complex energy  solutions of equation (\ref{EnEvenPE}). The blue and orange curves represent the solutions of the imaginary part of the equation for the plus and minus signs, respectively, and the colored points indicate the values of the complex energy that satisfy the full complex equation for selected values of the interaction strength $A_0$, shown in the legend. As the interaction strength $A_0$ varies, the solution to the full equation travels along the blue and orange curves. In particular, at a specific value of the interaction strength ($A_0=0.47$ in the plot), the solution reaches the supercritical energy $E=-m$, which corresponds to the absorption of a bound state. The bound state is eventually emitted at the critical energy $E=+m$ (corresponding to the value $A_0=8.5$ in the plot).

\begin{figure}[h!]
\includegraphics[scale=1.3]{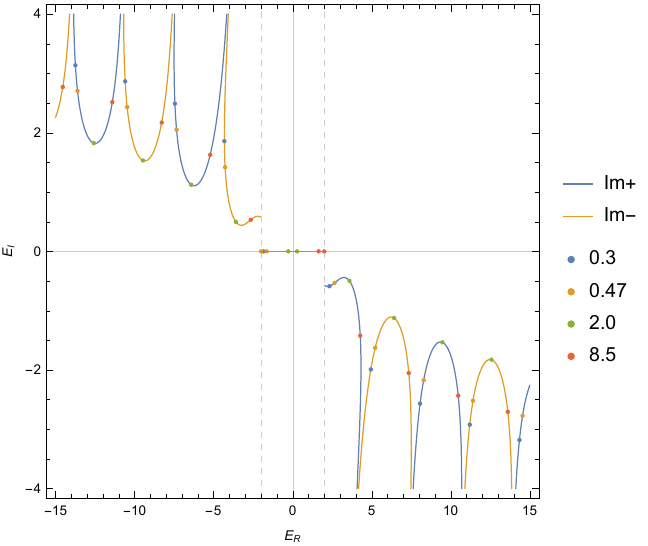}
\caption{ \label{figRES2PureElectrostaticEven} Complex energies corresponding to purely outgoing solutions of the Dirac equation with an even arrangement of two electrostatic interactions. In the figure $m=2$ and $\ell=1$. The blue and the orange curves correspond to complex energy solutions for the imaginary part of the complex equation (\ref{EnEvenPE}), with the plus and minus sign, respectively.  The colored points are the complex energies that solve the complex equation, for some values of the strength $A_0\in [0,+\infty)$ shown in the legend. In the plot, $A_0\approx 0.47$ corresponds to the {absorption} of a bound state at the supercritical energy $E=-m$, and $A_0\approx 8.5$ corresponds to its emission at $E=+m$.}
\end{figure}

\end{itemize}

%
%

\subsection{Odd Arrangements}

\subsubsection{Equal Mixtures of Scalar and Electrostatic Interactions}

An odd arrangement of two equal mixtures of electrostatic and scalar interactions ($A_0=B, A_1=W=0$) corresponds to take $a=d=1$, $\varphi=b=0$ and $c=2A_0=2B=A_0+B$ in (\ref{Lodd}), see Appendix \ref{app strengths}. Below we analyze, similarly to what we did for even arrangements, the critical, supercritical, bound and resonant states.

\begin{itemize}

\item \emph{Critical and supercritical states}. Condition (\ref{critodd}) implies $c=0$ ($A_0=B=0$) for critical states, and (\ref{superodd}) yields a supercritical state for \emph{every} value of $c$. This means that there is no emission/absorption of bound states at the supercritical energy. 

\item \emph{Bound states}. The condition for bound states (\ref{oddCond1}) may be rewritten as 
\begin{equation} \label{OddEqualMix}
\frac{(m+E)}{(m-E)} \left(1- e^{-2\kappa \ell}\right) = \frac{1}{A_0^2}\, ,
\end{equation}
which is symmetric under the change $A_0\to-A_0$. The system has a single bound state for \emph{any} value of the strength $A_0$, as shown in Figure \ref{figBS2deltasODD}; there is one particle bound state for any $A_0<0$ and one antiparticle bound state any $A_0>0$. It follows from (\ref{OddEqualMix}) that in the limit $\ell \to 0^+$ we obtain a free particle {since} the interactions cancel out. When $\ell \to \infty$, the bound state corresponds to that of a single equal mixture interaction, \textit{i.e.}, $E_b = \left(\frac{1-A_0^2}{1+A_0^2}\right)m$.

\comlm{It should be noted that our results for the bound states of both even and odd arrangements of equal mixtures of scalar and electrostatic interactions do not seem to coincide with those obtained \cite{GSZ22} for the same systems. It is likely that this discrepancy is a consequence of the regularization and implicit prescriptions used in \cite{GSZ22} to define the interactions. We note that our approach is rigorous and coincides with SAE where results from both approaches are available -- in fact, the non-relativistic limit of \eqref{BSEvenNLDeltaP} coincides with the rigorously obtained results of \cite{FGN24}. }

\begin{figure}[h!]
\includegraphics[scale=1]{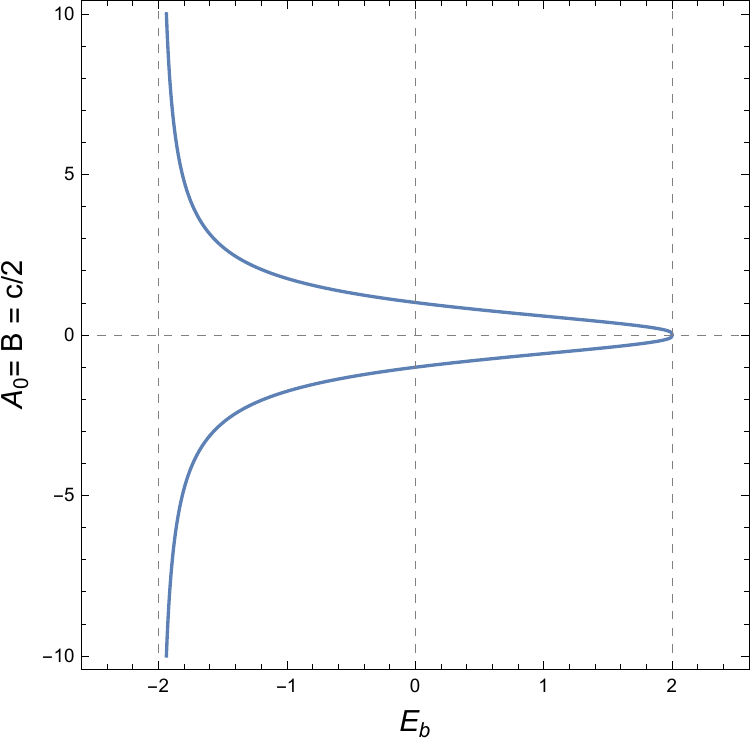}
\caption{\label{figBS2deltasODD}Bound states versus strength $A_0=B=\frac{c}{2}$ for an odd arrangement of two equal mixtures of scalar and electrostatic point interactions. There is a single bound state for any value  of the interaction strength, which corresponds to a particle's bound state for $A_0=B<0$ and to an antiparticle's one  for $A_0=B>0$.}
\end{figure}

\item \emph{Resonances}. The complex energies that solve equation (\ref{OddEqualMix}) with $\kappa\to -i k$ are shown in Figure \ref{figRES2deltasODD}. The  blue curve represents the solutions associated with the imaginary part of the equation, which is independent of the interaction strength $A_0$. The discrete colored points correspond to solutions that  satisfy the real and imaginary parts of equation \eqref{OddEqualMix} simultaneously for a selection of values of  $A_0$. As the interaction strength varies, the complex solutions move in the complex energy plane along the curve shown. For $|E_R|>m$, the solutions move toward the real axis as $A_0$ increases, while for $|E_R|<m$ they shift horizontally. The spectrum is symmetric under the transformation $A_0\to -A_0$. In the limit $A_0\to\infty$, the solutions become real and are given by the discrete energy levels of a particle or antiparticle confined between two impermeable walls located at $x=-\ell$ and $x=+\ell$.

\begin{figure}[h!]
\includegraphics[scale=1.3]{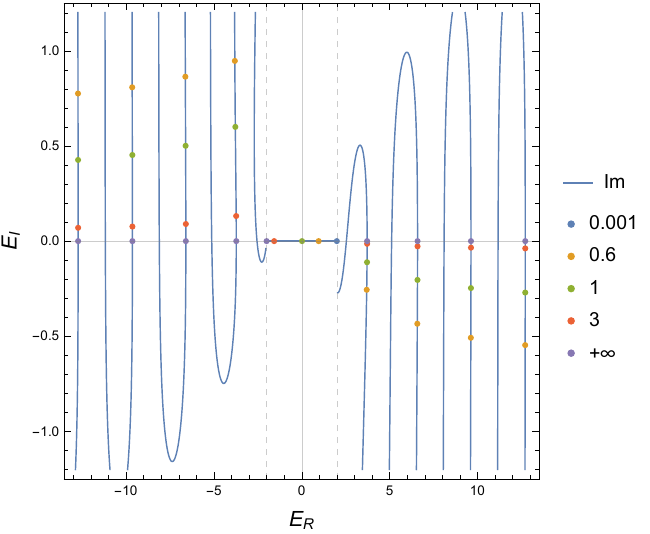}
\caption{\label{figRES2deltasODD} Complex solutions of equation (\ref{OddEqualMix}) for an odd arrangement of two equal mixtures of scalar and electrostatic interactions ($A_0=B$, $A_1=W=0$). The blue curve is the \emph{locus} of the complex energy solutions to the imaginary part of  the equation, which does not depend on $A_0$, whereas the colored points correspond to the solutions for both the real and imaginary parts of (\ref{OddEqualMix}), for several values of $A_0$ (indicated in the legend). The resonances satisfy the symmetry $A_0\to -A_0$. The points on the real axis are the allowed energies for a particle (or antiparticle) confined between two impermeable walls  at $x=-\ell$ and $x=+\ell$, and correspond to $A_0\to \infty$. As the interaction strength varies from $0$ to $+\infty$, the colored points move toward the real axis if $\left|E_R\right|>m$ and from right to left if $\left|E_R\right|<m $ (but never crossing the boundary at $E_R=-m$).}
\end{figure}

\end{itemize}

\subsubsection{Inverted Mixtures of Scalar and Electrostatic Interactions} 

Here, we consider an odd arrangement of inverted mixtures of electrostatic and scalar interactions, obtained by assuming $A_0=-B$, $A_1=W=0$ in (\ref{Bodd})-(\ref{Wodd}). In terms of the $\Lambda$-parameters (see Appendix \ref{app strengths}), it follows that $\varphi=c=0$ and $a=d=1$ and $\tfrac{b}{2}= A_0=-B$.  

\begin{itemize}

\item\emph{Critical and supercritical states}. From (\ref{critodd}) and (\ref{superodd}) there is a critical state for every value of the strength $b$, since $c=0$. On the other hand, a supercritical state exists only when $b=0$ ($A_0=-B=0$). Therefore, there is no absorption/emission of bound states at a critical state.

\item\emph{Bound states and resonances}. For this interaction, eq. (\ref{oddCond1}) can be conveniently written in terms of $A_0$ (keeping in mind that $\tfrac{b}{2}= A_0=-B$) as
\begin{equation} \label{InvOdd}
\left(\frac{m-E}{m+E}\right) (1- e^{-2\kappa \ell}) = \frac{1}{A_0^2}\,,
\end{equation}
which can be promptly obtained from (\ref{OddEqualMix}) under the substitution $E\to -E$ (equivalently, $m\to -m$). Therefore, the entire structure of states for this interaction can inferred from the results of the previous subsection.  Recall that such a symmetry between the equal and inverted mixtures of electrostatic and scalar interactions was also observed in the case of even arrangements. 

\end{itemize}

\subsubsection{Two Pseudoscalar Interactions} 

Let us now consider an odd arrangement of two pseudoscalar point interactions, that is, $W$ is arbitrary and $B=A_0=A_1=0$ in (\ref{Bodd})-(\ref{Wodd}), which corresponds to $\varphi=b=c=0$ and $d=\frac{1}{a}=\frac{2+W}{2-W}$ in (\ref{Lodd}), according to the formulae in the Appendix \ref{app strengths}. A single pseudoscalar interaction is an odd interaction \cite{BLM24}. Thus, this system is an odd arrangement of two single odd interactions and, as noted before [see comments after the equation (\ref{OddCond2})], this kind of system has no bound states.

\begin{itemize}

\item\emph{Critical and supercritical states.} Equations (\ref{critodd}) and (\ref{superodd}) gives critical and supercritical states for any value of the strength parameter $W$. Therefore, there is no emission/absortion of bound states for this system.

\item \emph{Bound states.} Equation (\ref{oddCond1}) reduces to
$$
\left(a-\frac{1}{a}\right) e^{-\kappa \ell}=\pm \sqrt{-\left(a+\frac{1}{a}\right)^2}\, . 
$$
\comjt{The lhs of the above equation is a real number, whereas the rhs is purely imaginary, since $a$ is real and $\kappa>0$. This equation has no solution for real $E$ and, thus, } there is no bound state in this case.

\item\emph{Resonances.} For the parameters above, with $\kappa \to -i k \ell$, condition (\ref{oddCond1}) can be written as 
\begin{equation}\label{PPSodd}
e^{-2 ik \ell}=-\left(\frac{a-d}{a+d}\right)^2=-\left(\frac{4W}{4+W^2}\right)^2\; .
\end{equation}
This equation is symmetric under $W\to -W$ and $W\to\frac{4}{W}$; thus, it suffices to investigate the structure of resonances for $W\in[0,2]$. In particular, the case $W=2$ corresponds to two impenetrable walls at $x=-\ell$ and $x=+\ell$. The imaginary part of (\ref{PPSodd}) is the same as for the even arrangement of pseudoscalar interactions [compare with eq. (\ref{resPS})]. Therefore, the even and odd arrangements will differ only by the solutions of the real part of the corresponding equations. Figure \ref{figRES2localdeltasODD} shows the resonances for several values of the strength $W\in[0,2]$. The structure of the resonances is very similar to the one for the even arrangement of two pseudoscalar interactions; the difference is that now the real part of the complex energies are located on the complementary set of blue curves with respect to those in the even arrangement case. As a result, the real energies are approximately translated in the real axis.
\begin{figure}[h!]
\includegraphics[scale=1.3]{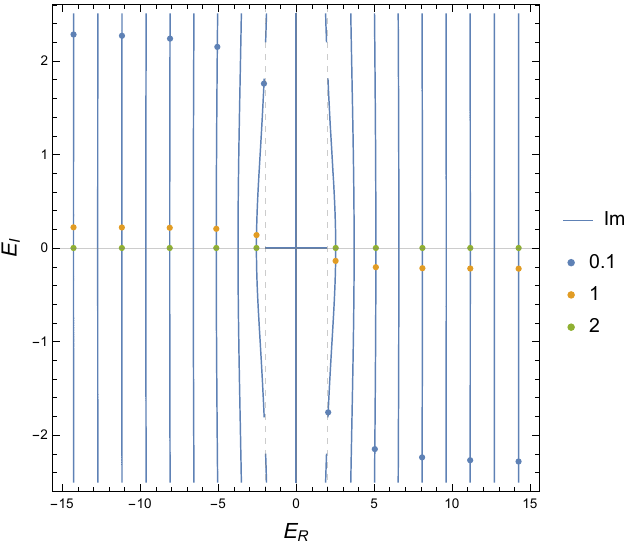}
\caption{\label{figRES2localdeltasODD}{Complex solutions of equation (\ref{PPSodd}) for an odd arrangement of two pseudoscalar point interactions. The blue curve is the \emph{locus} of the complex energy solutions to the imaginary part of  the equation (\ref{PPSodd}), which does not depend on the strength $W$, whereas the colored points correspond to the solutions of both the real and imaginary parts, for three values (indicated in the legends) of the strength in the interval $W\in[0,2]$. The points on the real axis are the allowed energies for a particle (or antiparticle) confined between two impermeable walls  at $x=-\ell$ and $x=+\ell$, and correspond to $W=2$.}}
\end{figure}

\end{itemize}

\subsubsection{Two Magnetostatic Point Interactions} 

Similarly to the case of even arrangements, also  for odd arrangements the two magnetostatic interactions do not posses any structure of bound states, critical/supercritical states, or resonances, since only the phase $\varphi$ is nonvanishing in (\ref{Lodd}). Thus, the system behaves essentially as a free particle. 


\subsubsection{Two Scalar Point Interactions}

In this case the following parameters characterize the odd arrangement (\ref{Lodd}): $B$ arbitrary, $W=A_0=A_1=0$. From the expressions in Appendix \ref{app strengths}, the $\Lambda$-matrix parameters become $d=a=\frac{4+B^2}{4-B^2}$, $b=c=\frac{4B}{4-B^2}$, and it follows that for $|B|=2$ the two point barriers are impenetrable.

Note that the transformation $B\to\frac{4}{B}$ has the effect of multiplying all the parameters $a,b,c,d$ by $-1$, which corresponds to multiplying the $\Lambda$-matrices in (\ref{Lodd}) by a phase factor. Naturally, such a phase factor has no effect on the conditions for the existence of physical states. Thus, it suffices to investigate the strength parameter $B$ in the interval $B\in [-2,2]$.

\begin{itemize}

\item\emph{Critical and supercritical states}. Conditions (\ref{critodd}) and (\ref{superodd}) for the existence of critical and supercritical states, respectively, can only be satisfied for $B=0$. 

\item\emph{Bound states}. Equation (\ref{oddCond1}) in this case may be written as
\begin{equation}\label{PSOdd}
 \frac{m^2-E^2\, e^{-2 \kappa \ell}}{\kappa^2}= \left(\frac{4+B^2}{4B} \right)^2\, ,
\end{equation}
which shows the additional symmetry $B\to -B$. Figure \ref{figBSPSOdd} shows the relationships between the bound state energies $E_b$ and the interaction strength $B$. We note that for $B\neq \pm 2$ there are always two bound states, with symmetric energies, for any  value of the strength $B$. For $B=\pm 2$, these bound-state energies coalesce to zero energy, which is one of the allowed energies for a particle (or antiparticle) confined between the two impermeable walls at $x=-\ell$ and $x=+\ell$. For $B=0$, there are a critical and a superctitical state, as seen above.
\begin{figure}[h!]
\includegraphics[scale=1]{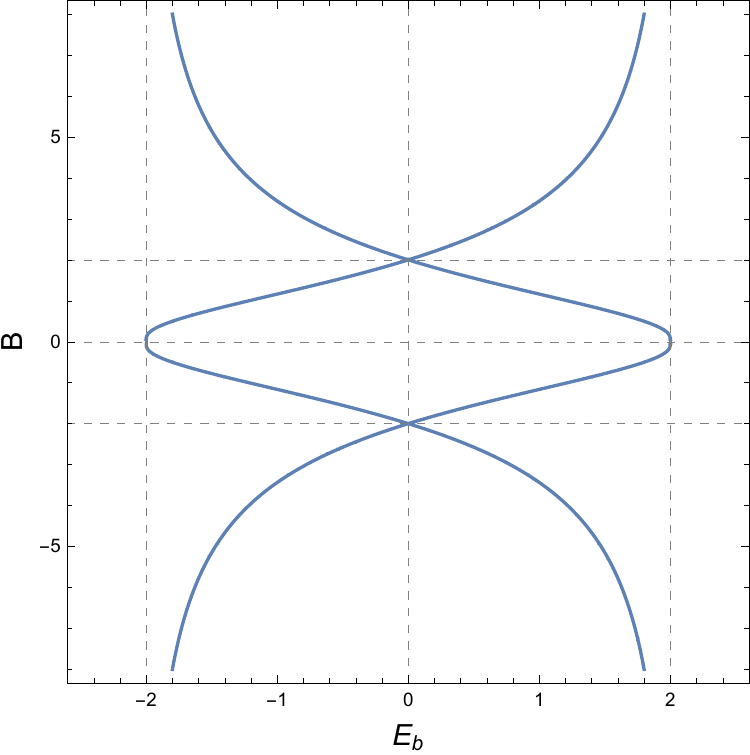}
\caption{\label{figBSPSOdd}Bound-state energies versus strength $B$ for an odd arrangement of two scalar point interactions. For any value (positive or negative) of $B\neq \pm 2$ there are two bound states. When $|B|=2$ the double barrier becomes impermeable, and the two bound states coalesce at zero energy, which is one member of the discrete set of allowed energies for particles (or antiparticles) confined between the two barriers.}
\end{figure}

\item\emph{Resonances}. The complex energy solutions of equation (\ref{PSOdd}), with $\kappa\to -i k$, are shown in Figure \ref{figRESPSOdd} for some values of the strength parameter $B\in [0,2]$ -- note that \eqref{PSOdd} implies the symmetry $E\to -E$.  From Figure \ref{figRESPSOdd} we observe that there are neither emissions or absorptions of bound states at the critical and supercritical energies when $B=0$. From the symmetry $B\to -B$ we observe that as the interaction strength $B$ crosses the zero value, the bound-state energies are ``reflected" at $E_b\pm m$. As $B$ increases in the interval $[0,2]$, the complex energies approach the real axis; for $B=2$ the two point barriers become impenetrable and the energies form the set of discrete, \emph{real} energies allowed for a particle (or antiparticle) confined between the two impenetrable walls at $x=-\ell$ and $x=+\ell$. Similar conclusions for other values of $B$ can be drawn from the symmetries $B\to \frac{4}{B}$ and $B\to -B$.
\begin{figure}[h!]
\includegraphics[scale=1.3]{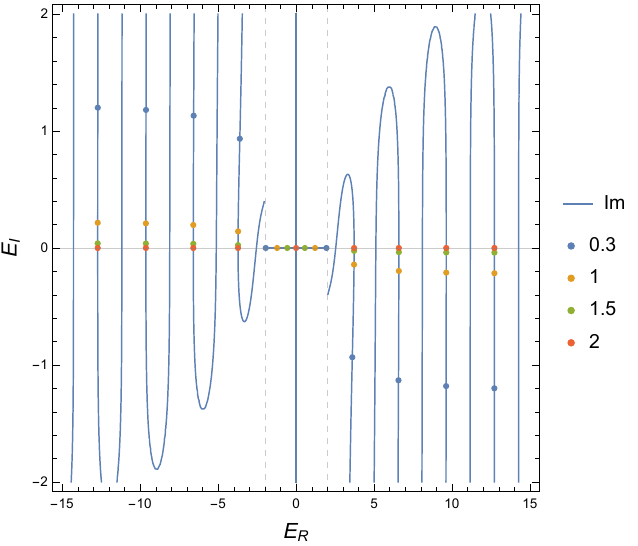}
\caption{\label{figRESPSOdd}{Complex solutions of equation (\ref{PSOdd}) for an odd arrangement of two pseudoscalar point interactions. The blue curve is the \emph{locus} of the complex energy solutions to the imaginary part of  the equation, which does not depend on the strength $W$, whereas the colored points correspond to the solutions for both the real and imaginary parts of it, for three values (indicated in the legends) of the strength in the interval $W\in[0,2]$. The resonances satisfy the symmetries $W\to -W$ and $W\to \frac{4}{W}$. The points on the real axis are the discrete energies allowable to a particle (or antiparticle) confined between two impermeable walls  at $x=-\ell$ and $x=+\ell$, and correspond to $|W|=2$.}}
\end{figure}

\end{itemize}


\subsubsection{Two Electrostatic Point Interactions}

As the final case, we now consider an odd arrangement of two electrostatic point interactions, that is, $A_0$ arbitrary and $A_1=B=W=0$. It follows from Appendix \ref{app strengths} that this choice of parameters corresponds to $a=d=\frac{4-A_0^2}{4+A_0^2}$ and $c=-b=\frac{4A_0}{4+A_0^2}$ in (\ref{Lodd}). As already mentioned when we considered the even arrangements in subsection \ref{evensubsec}, the two single electrostatic barriers are never impenetrable, since the parameters $a,b,c,d$ are all finite (in terms of the interaction strengths, the permeability condition (\ref{permcond}) is satisfied).

\begin{itemize}

\item\emph{Critical and supercritical states}. Both (\ref{criteven}) and (\ref{critodd}) give the condition $c=0$, which is equivalent to $A_0=0$, for the existence of both critical and supercritical states. 

\item \emph{Bound states}. Substituting the above parameters in (\ref{oddCond1}), the condition for bound states becomes
\begin{eqnarray}\nonumber
\frac{E^2-m^2 e^{-2 \kappa \ell}}{\kappa^2}&=&\left(\frac{a}{b}\right)^2\\
&=& \left(\frac{4-A_0^2}{4A_0}\right)^2\, . \label{OddPE}
\end{eqnarray}
This equation is symmetric under each of the transformations $A_0\to\frac{4}{A_0}$ and $A_0\to -A_0$. Figure \ref{figBSPEodd} shows the relationship between the bound-state energies and the interaction strength $A_0$. We observe that there are always two bound states, independently of the value of $A_0$. We also observe that the energies of these two bound states have the symmetry $E\to -E$, as is evident from (\ref{OddPE}). This figure also shows that as $A_0$ crosses the zero value there is neither emission nor absorption of bound states to or from the continuum, but only a ``reflection" of the bound states at $E=\pm m$. 
\begin{figure}[h!]
\includegraphics[scale=1]{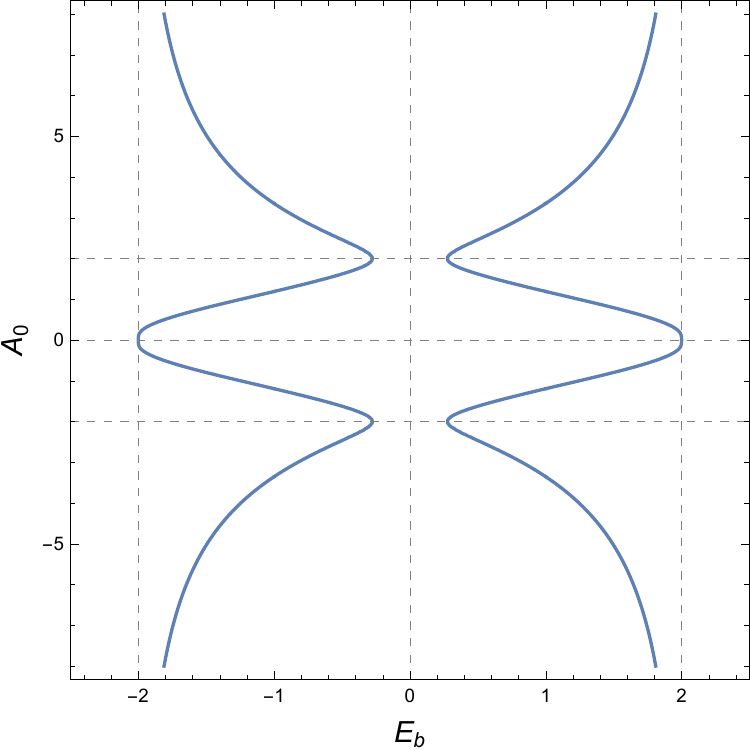}
\caption{\label{figBSPEodd}Bound states versus strength $A_0$ for an odd arrangement of two electrostatic point interactions. For any value of $A_0$ there are two bound states, with symmetric energies. These kind of point barriers are never impenetrable, and the odd arrangement has the symmetries $A_0\to \frac{4}{A_0}$ and $A_0\to -A_0$.}
\end{figure}

\item\emph{Resonances}. Now we seek for the complex energy solutions of (\ref{OddPE}), with $\kappa \to -i k$. Similarly to the previous cases, only the real part of that equation depends on the interaction strength $A_0$. By taking into account the symmetries $A_0\to \frac{4}{A_0}$ and $A_0\to -A_0$ it is sufficient to investigate the resonances with the interaction strength in the interval $A_0\in [0,2]$. Figure \ref{figRESPEodd} shows the structure of the resonances, for three values of $A_0\in [0,2]$. As a consequence of the fact that electrostatic barriers are never impenetrable and thus are unable to confine a particle (or antiparticle) in the region between the point barriers, the resonances can never be real, regardless of the value of $A_0$. As $A_0$ increases over the interval $[0,2]$ the complex energy solutions (colored points in Figure \ref{figRESPEodd}) move along the blue curve towards the real axis, but never reach it since when $A_0$ crosses the value $A_0=2$ the colored points in the figure revert their movement along the blue curve. The behavior of the resonances for $A_0$ outside the interval $[0,2]$ can be inferred from the symmetries $A_0\to \frac{4}{A_0}$ and $A_0\to -A_0$. Finally, the bound states are also represented in Figure \ref{figRESPEodd} as colored points within the real interval $[-m,+m]$. As $A_0$ increases within the interval $A_0\in [0,2]$ the bound-state energies move towards the origin, but never reach it; as $A_0$ crosses the value $2$ the colored points reverse their motion within the bound-state interval $[-m,+m]$.
\begin{figure}[h!]
\includegraphics[scale=1.5]{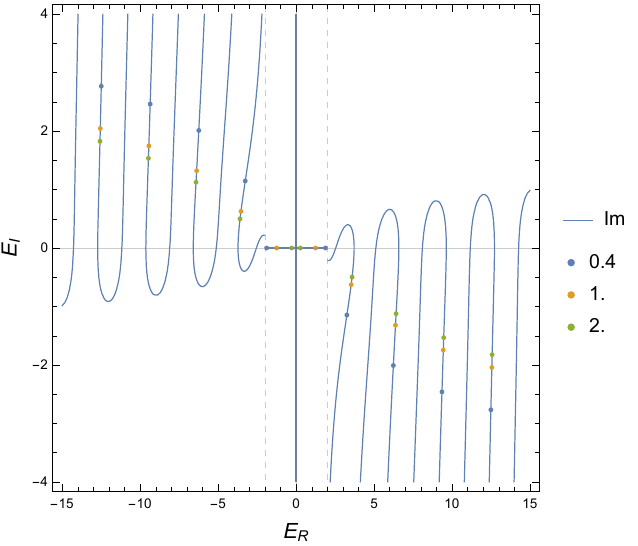}
\caption{\label{figRESPEodd}{Complex solutions of equation (\ref{OddPE}) for an odd arrangement of two electrostatic point interactions. The blue curve is the \emph{locus} of the complex energy solutions to the imaginary part of  the equation, which does not depend on the strength $A_0$, whereas the colored points correspond to the solutions for both the real and imaginary parts of it, for three values of the strength $A_0$ in the interval $[0,2]$. The resonances for $A_0$ outside of this interval can be obtained from the symmetries $A_0\to \frac{4}{A_0}$ and $A_0\to -A_0$.}}
\end{figure}

\end{itemize}

\subsection{Results summary}
\comcb{
To end this section, we compile the results for the particular arrangements studied in two tables. Table \ref{tab: even} summarizes the results obtained for even arrangements and Table \ref{tab: odd} does the same for the odd ones.}

\begin{table}[htbp]
\centering

\setlength{\tabcolsep}{4pt}
\renewcommand{\arraystretch}{1.1}
{\tiny
\begin{NiceTabular}{p{2.3cm} p{2.3cm} p{2.3cm} p{2.3cm} p{2.3cm} p{2.3cm}}[hvlines]
\textbf{Interaction type } &
\textbf{Parameters } &
\textbf{Critical States} ($E=+m$) &
\textbf{Supercritical States} ($E=-m$) &
\textbf{Bound States} ($|E|<m$) &
\textbf{Resonances} ($E\notin\mathbb{R}$) \\
Scalar + electrostatic: two equal mixtures & $A_0=B$ arbitrary, $A_1=W=0$  & $A_0=0$ or $A_0=-1/(2m\ell)$; bound states absorbed/emitted at these values & For all $A_0$; no bound state absorption/emission & Only for $A_0<0$. As $A_0$ decreases, the ground state appears at $A_0=0$; excited state appears at $A_0=-1/(2m\ell)$ & Yes; become real at $|A_0|\to\infty$ (impenetrable barriers) \\
Scalar + electrostatic: two inverted mixtures  & $A_0=-B$ arbitrary, $A_1=W=0$  & For all $A_0$; no bound state absorption/emission & $A_0=0$ or $A_0=-1/(2m\ell)$; bound states absorbed/emitted at these values & Same  as two equal mixtures under $E\to -E$  & Same as two equal mixtures under $E\to -E$\\
Two pseudoscalar & $W$ arbitrary, $A_0=A_1=B=0$ & For all $W$; no bound state absorption/emission & For all $W$; no bound state absorption/emission & None & Yes; become real at $|W|=2$ (impenetrable barriers)\\
Two magnetostatic & $A_1$ arbitrary, $A_0=B=W=0$ & None (free-like) & None (free-like) & None (free-like) & None (free-like)\\
Two scalar & $B$ arbitrary, $A_0=A_1=W=0$ & $B=0$ or $B_\pm = -2m\ell \mp 2\sqrt{(m\ell)^2-1}$; $B_{\pm}$ real iff $\ell\geq 1/m$; bound states absorption/emission at these values & $B=0$ or $B_\pm = -2m\ell \mp 2\sqrt{(m\ell)^2-1}$; $B_{\pm}$ real iff $\ell\geq 1/m$; bound states absorption/emission at these values & At least two for $B< 0$; additional pair if $B\in(B_+,B_-)$ when $\ell\geq 1/m$ & Yes; become real at $|B|=2$ (impenetrable barriers) \\
Two electrostatic & $A_0$ arbitrary; $A_1=B=W=0$ & $A_0=0$ or $A_{0\pm}^{\mbox{crit}}=2m\ell\pm 2\sqrt{(m\ell)^2+1} $ with $|A_{0\pm}^{\mbox{crit}}| \neq 2$; bound states absorption/emission at these values;  & $A_0=0$ or $A_{0\pm}^{\mbox{super}}=-2m\ell\pm 2\sqrt{(m\ell)^2+1}$ with $|A_{0\pm}^{\mbox{super}}|\neq 2$; bound states absorption/emission at these values& At least one for any $A_0$; number changes as $A_0$ crosses critical/supercritical values & Yes; never become real (barriers always penetrable)

\end{NiceTabular}
}
\caption{Existence (or not) of Resonances and of Critical, Supercritical, and Bound States for even arrangements for all cases studied in the text. \label{tab: even}}
\end{table}

\begin{table}[htbp]
\centering

\setlength{\tabcolsep}{4pt}
\renewcommand{\arraystretch}{1.1}
{\tiny
\begin{NiceTabular}{p{2.3cm} p{2.3cm} p{2.3cm} p{2.3cm} p{2.3cm} p{2.3cm}}[hvlines]
\textbf{Interaction type} &
\textbf{Parameters} &
\textbf{Critical States} ($E=+m$) &
\textbf{Supercritical States} ($E=-m$) &
\textbf{Bound States} ($|E|<m$) &
\textbf{Resonances} ($E\notin\mathbb{R}$) \\
Scalar + electrostatic: two equal mixtures & $A_0=B$ arbitrary, $A_1=W=0$ & Only for $A_0=0$  & For all $A_0$; no bound state absorption/emission & Exactly one for any $A_0$: particle bound state for $A_0<0$, antiparticle for $A_0>0$ & Yes; become real as $|A_0|\to\infty$ (impenetrable barriers)\\
Scalar + electrostatic: two inverted mixtures & $A_0=-B$ arbitrary, $A_1=W=0$ & For all $A_0$; no emission/absorption & Only for $A_0=0$ & Same  as two equal mixtures under $E\to -E$ & Same  as two equal mixtures under $E\to -E$ \\
Two pseudoscalar & $W$ arbitrary, $A_0=A_1=B=0$ & For all $W$: no bound state absorption/emission & For all $W$; no bound state absorption/emission & None & Yes; become real at $|W|=2$ (impenetrable barriers)\\
Two magnetostatic & $A_1$ arbitrary, $A_0=B=W=0$ &  None (free-like) & None (free-like) & None (free-like) & None (free-like)\\
Two scalar & $B$ arbitrary, $A_0=A_1=W=0$ &  Only for $B=0$ & Only for $B=0$ & Two bound states for any $|B|\neq 2$ with $E\to - E$ symmetry; for $|B|=2$ they coalesce at $E=0$ & Yes; become real at $|B|=2$ (impenetrable barriers)  \\
Two electrostatic & $A_0$ arbitrary; $A_1=B=W=0$ & Only for $A_0=0$ & Only for $A_0=0$ & Always two bound states with $E\to -E$ symmetry & Yes; never become real (barriers always penetrable)

\end{NiceTabular}
}
\caption{Existence (or not) of Resonances and of Critical, Supercritical, and Bound States for odd arrangements of interaction for all cases studied in the text. \label{tab: odd}}
\end{table}

\section{Concluding remarks} 
\label{conclusion}

We have investigated a relativistic, one-dimensional model with two contact interactions located symmetrically with respect to the origin, using the mathematically rigorous distributional method developed by some of the authors. Taking advantage of this approach, each of the contact interactions depends on four independent parameters with a well-defined physical meaning: scalar, electrostatic, magnetic, and pseudoscalar interactions. These interactions can be equivalently characterized by some matching conditions for the wave function at the points supporting the interaction. We discuss the form of the coefficients and matching conditions for two particular cases of special interest: even and odd interactions. We have also considered the limit of one barrier, in which both barriers are supported at the same point, for both even and odd barriers. Some interesting results have emerged after this limit: while the one-point limit of an even arrangement is always an even interaction, the $\ell\to 0^+$ limit of an odd arrangement does not always have a well-defined symmetry.

Using the form of the wave function outside the interaction points and the matching conditions at these points, we constructed the transmission matrix and, hence, the scattering matrix. Then, we provided a detailed analysis of the structure of critical, supercritical, and bound states, as well as of the resonances for even and  odd arrangements of interactions for all the main relativistic, point interactions of physical interest in the literature, namely, equal and inverted mixtures of electrostatic and scalar interactions, pseudoscalar, scalar, magnetostatic and electrostatic interactions. 

A possible extension of this work would be to consider interactions with $N$ contact interactions, i.e., interaction supported at $N$ points, and investigate the transmission bands. \comlm{This requires the construction of the transfer matrix for the system of $N$ interactions, for an even number of interactions the results obtained here can be directly applied, since the comb can be seen as $N/2$ cells of two interactions. This will provide reflection and transmission coefficients and, hence, bound states and scattering features such as resonances.}  Additionally, as was demonstrated, in order to confine a particle or an anti-particle on one side of the interaction (\textit{i.e.}, to have an impermeable interaction), the single-point interactions \emph{must} include a scalar and/or a pseudoscalar component, since electrostatic interactions are non-confining (see Figures \ref{figRES2PureElectrostaticEven} and \ref{figRESPEodd}). This could be related to the Klein effect and deserves further study.


\appendix

\section{The Strengths of the Interactions}
\label{app strengths}

Here we briefly reproduce the relationships between the physical parameters (potential strengths) and the parameters of the $\Lambda$-matrix -- for details on the derivation of these results, we refer the reader to \cite{BLM24} (see also \cite{HTu22, BHT23}). At each point supporting the interaction, the physical parameters can be obtained from the elements of each $\Lambda$-matrix in \eqref{lbda} through: 

\begin{eqnarray}
B&=& \frac{2(c+b)}{2\cos\varphi+d+a} \label{Bpar}\, , \label{B}\\
A_0&=& \frac{2(c-b)}{2\cos\varphi+d+a}\, , \label{A0}\\
A_1&=& \frac{-4\sin\varphi}{2\cos\varphi+d+a}\, ,\label{A1}\\
W&=& \frac{2(d-a)}{2\cos\varphi+d+a} \label{W}
\, .
\end{eqnarray}

The inverse relationships are given by
\begin{eqnarray}\label{fp}
\varphi&=&\tan^{-1}\left(\frac{4 A_1}{B^2+W^2-4-A_0^2+A_1^2}\right)\,\qquad \varphi\in [0,\pi)\, ,\\ \nonumber \\ \label{ap}
a&=& \eta \,\frac{A_0^2-A_1^2-B^2
   -(W-2)^2}{\sqrt{\left(B^2+W^2-4-A_0^2+A_1^2\right)^2+16A_1^2}}\, ,\\ \label{bp}
b&=& \eta\, \frac{4 (A_0-B)}{\sqrt{\left(B^2+W^2-4-A_0^2+A_1^2\right)^2+16A_1^2}}\, ,\\ \nonumber \\ \label{cp}
c&=& \eta \,\frac{-4 (A_0+B)}{\sqrt{\left(B^2+W^2-4-A_0^2+A_1^2\right)^2+16A_1^2}}\, ,\\ \nonumber \\ \label{dp}
d&=& \eta\, \frac{A_0^2-A_1^2-B^2
   -(W+2)^2}{\sqrt{\left(B^2+W^2-4-A_0^2+A_1^2\right)^2+16A_1^2}}\, ,
\end{eqnarray}
\comjt{where $\eta=\mathrm{sign}\, A_1$, if $A_1\neq 0$ and  $\eta=\mathrm{sign}\,\left(B^ 2+W^ 2-4-A_0^ 2\right)$ otherwise.}

\comjt{Above, we may consider the $\mathbb{\Lambda}$-matrix parameters as being all finite or (some of them) infinite. The point interaction is said to be \emph{permeable} (or \emph{penetrable}) if the  $\mathbb{\Lambda}$-matrix parameters are all finite; in this case equation (\ref{bc}) completely determines $\psi(x_j^+)$ from the knowledge of $\psi(x_j^-)$, and vice-versa. If any of the $\mathbb{\Lambda}$-matrix parameters were infinite, the point interaction is said to be \emph{impermeable} (or \emph{impenetrable}); in this case $\psi(x_j^+)$ \emph{is not} completely determined from $\psi(x_j^-)$, and the solutions in both sides of the origin are independent from each other (see \cite{BLM24} for details). The physical strengths $B,A_{\mu}$ and $W$ may either be all finite or (at least some of them) infinite; in any case, the permeability condition implies that} 
\begin{equation}\label{permcond}
    A_1\neq 0 \qquad\text{or}\qquad B^2+W^2-A_0^2-4\neq 0\, .
\end{equation}
\comjt{If the physical parameters are all finite, condition (\ref{permcond}) is \emph{equivalent} to the permeability condition. Finally, if any the physical strengths are infinite and the point interaction is impermeable, condition (\ref{permcond}) may or may not be satisfied. For example, the interaction with $B=A_0, \,A_{1}=W=0$ (which corresponds to the ``delta interaction" in non-relativistic limit), with $A_0$ (and $B$) infinite, is impermeable at the singular point, but the physical parameters satisfy (\ref{permcond}). On the other hand, the interaction with $B=A_0, \,A_{1}=0$, but $W=2$ is impermeable and violates (\ref{permcond}), regardless of whether $B=A_0$ is finite or infinite.}



\section{On the non-relativistic limit}
\label{app NR}

In the present appendix, and for the sake of completeness, we study the non-relativistic limit of the Dirac equation \eqref{dir} with \eqref{d}. As a matter of fact, this task has been performed in \cite{BLM24}, so that we here just recall the results in that paper adapted to our particular situation. 

As is well known, the solution of the Dirac equation \eqref{dir} is a two component spinor of the form $\psi(x)=(u(x),v(x))^T$, where the superindex $T$ denotes the transpose. Then, equation \eqref{dir} with \eqref{d} may be split into two equations both in $u(x)$ and $v(x)$ in the following form (the prime means derivative with respect to the variable $x$:

\begin{eqnarray}\label{50}
&&(E-m) u(x) +i v'(x)= \nonumber \\[2ex] &&\qquad= \left[ (B^{(1)} + A_0^{(1)}) \left( \frac{u_-^+ + u_-^-}{2} \right) + (iW^{(1)} + A_1^{(1)}) \left( \frac{v_-^+ + v_-^-}{2} \right) \right] \delta (x-\ell/2) \nonumber \\[2ex] &&\qquad+ \left[ (B^{(2)} + A_0^{(2)}) \left( \frac{u_+^+ + u_+^-}{2} \right) + (iW^{(2)} + A_1^{(2)}) \left( \frac{v_+^+ + v_+^-}{2} \right) \right] \delta(x+ \ell/2)\,,
\end{eqnarray}
and
\begin{eqnarray}\label{51}
&&-(E+m)v(x) -iu'(x) =\nonumber \\[2ex] &&\qquad= \left[ (iW^{(1)} -A_1^{(1)}) \left( \frac{u_-^+ + u_-^-}{2} \right) + (B^{(1)} -A_0^{(1)}) \left( \frac{v_-^+ + v_-^-}{2} \right) \right] \delta (x-\ell/2) \nonumber \\[2ex] &&\qquad+ \left[ (iW^{(2)} -A_1^{(2)}) \left( \frac{u_+^+ + u_+^-}{2} \right) + (B^{(2)} -A_0^{(2)}) \left( \frac{v_+^+ + v_+^-}{2} \right) \right] \delta (x+\ell/2)\,,
\end{eqnarray}
where $u_-^+:= u(-\ell^+/2)$, $u_-^- := u(-\ell^-/2))$, $u_+^+:= u(\ell^+/2)$, $u_+^-:= u(\ell^-/2)$ and the same notation for $v$ instead of $u$. We recall that $u(\alpha^\pm):= \lim_{\varepsilon \to 0} u(\alpha \pm \varepsilon)$.

In order to have the non-relativistic limit, we first note that 

\begin{equation}\label{52}
E= m+\varepsilon\,, \qquad E+m \approx 2m\,,
\end{equation}
where $\varepsilon$ is the non-relativistic energy. Outside the origin, the spinor $\psi(x)$ satisfies the free equation and its ``small'' component $v(x)$ is proportional to the ``large'' component $u(x)$ as may be seen from \eqref{51}, so that

\begin{equation}\label{53}
v(x)= -\frac{i}{2m}\, u'(x) \Longrightarrow v^\pm_\pm = -\frac{i}{2m}\, u'^\pm_\pm \,.
\end{equation}

Then, take the expression of $v(x)$ obtained from \eqref{51} in terms of $u'(x)$ and all the other terms, use this expression in \eqref{50} and take the non-relativistic limit using \eqref{52}. We obtain the following equation:

\begin{eqnarray}\label{54}
-\frac{1}{2m}\, u''(x) - \varepsilon\, u(x) \nonumber \\[2ex] = \left[ -(B^{(1)} + A_0^{(1)}) \left( \frac{u_-^+ + u_-^-}{2} \right) - \frac{W^{(1)} -i A_1^{(1)}}{2m} \left( \frac{u_-'^+ + u_-'^-}{2} \right)  \right] \delta(x-\ell/2) \nonumber \\[2ex] + \left[ \left( \frac{W^{(1)} + iA_1^{(1)}}{2m} \right) \left( \frac{u_-^+ + u_-^-}{2} \right)  - \left( \frac{B^{(1)} -A_0^{(1)}}{4m^2}\right) \left( \frac{u_-'^+ + u_-'^-}{2}  \right)\right] \delta'(x-\ell/2) +T\,,
\end{eqnarray}
where $T$ is a term that is similar to the two terms on the right hand side of \eqref{54} with the following modifications: i) the superscript ``$(1)$'' has to be replaced by the superscript ``$(2)$''; ii) We have to perform the following replacements: $u_-^\pm \longmapsto u_+^\pm$ and $u_-'^\pm \longmapsto u_+'^\pm $ and iii) replace $-\ell/2 \longmapsto +\ell/2$. In addition, $\delta'(x \pm \ell/2)$ is the distributional derivative of $\delta(x \pm \ell/2)$. 

Now, we discuss some particular cases that we consider as of special interest:

\begin{itemize}

\item{{\bf Delta interactions}. 

From \eqref{54}, a delta interaction appears at the point $\pm \ell/2$ only when ($i=1,2$):

\begin{equation}\label{55}
W^{(i)}=A_1^{(i)}=0\,, \qquad B^{(i)}=A_0^{(i)} \ne 0\,.
\end{equation}

This interaction consists of an equal mixture of electrical and scalar point interactions. In this case, the Schr\"odinger equation reads as

\begin{equation}\label{56}
-\frac{1}{2m}\, u''(x) - \varepsilon\, u(x) = -2B^{(1)}\, \left( \frac{u_-^+ + u_-^-}{2} \right) \delta(x-\ell/2) -2B^{(2)}\, \left( \frac{u_+^+ + u_+^-}{2} \right) \delta(x+\ell/2)\,.
\end{equation}

In the non-relativistic case, we can define for each point $\pm \ell/2$ a matrix similar to \eqref{lbda}, which in the present case takes the form:

\begin{equation}\label{57}
\Lambda_i = \left(\begin{array}{cc} 1 & 0 \\[2ex] 4m B^{(i)} & 1  \end{array}  \right)\,, \qquad i=1,2\,.
\end{equation}
}

\item{{\bf Non-local $\delta'$ interactions.}

At each point, this interaction appears as an inverted mixture of electrostatic and scalar point interactions defined as ($i=1,2,$):

\begin{equation}\label{58}
W^{(i)}=A_1^{(i)} =0\,, \qquad B^{(i)}=-A_0^{(i)} \ne 0\,.
\end{equation}

The Schr\"odinger equation now reads:

\begin{equation}\label{59}
-\frac{1}{2m}\, u''(x) - \varepsilon\, u(x) = - \frac{2B^{(1)}}{4m^2} \, \frac{u_-'^+ + u_-'^-}{2} \, \delta'(x-\ell/2) + - \frac{2B^{(2)}}{4m^2} \, \frac{u_+'^+ + u_+'^-}{2} \, \delta'(x+\ell/2)\,,
\end{equation}
and

\begin{equation}\label{60}
\Lambda_i = \left( \begin{array}{cc} 1 & \displaystyle \frac{2B^{(i)}}{2m} \\[2ex] 0 & 1  \end{array} \right)\,, \qquad i=1,2\,.
\end{equation}
}

\item{{\bf Local $\delta'$ interactions.}

These interactions are produced by a pure pseudo-scalar interaction at each point. It corresponds to the choice ($i=1,2$):

\begin{equation}\label{61}
B^{(i)}=A_0^{(i)}=A_1^{(i)}=0\,, \qquad W^{(i)} \ne 0\,.
\end{equation}

The Schr\"odinger equation is

\begin{eqnarray}\label{62}
-\frac{1}{2m}\, u''(x) - \varepsilon\, u(x) =  \frac{W^{(1)}}{2m} \left[- \frac{u_-'^+ + u_-'^-}{2}\, \delta(x-\ell/2) + \frac{u_-^+ + u_-^-}{2} \, \delta'(x-\ell/2) \right] \nonumber \\[2ex] + \frac{W^{(2)}}{2m} \left[- \frac{u_+'^+ + u_+'^-}{2}\, \delta(x+\ell/2) + \frac{u_+^+ + u_+^-}{2} \, \delta'(x+\ell/2) \right].
\end{eqnarray}

At each point, we have the following matching conditions:

\begin{equation}\label{63}
\Lambda_i = \left( \begin{array}{cc} \displaystyle\frac{2-W^{(i)}}{2+W^{(i)}} & 0 \\[2ex] 0 & \displaystyle\frac{2+W^{(i)}}{2-W^{(i)}} \end{array} \right)\,, \qquad W^{(i)} \ne 2\,, \qquad i=1,2\,.
\end{equation}
}

\item{{\bf The {\it singular gauge field} interaction.}

Here,

\begin{equation}\label{64}
B^{(i)} = A_0^{(i)}= W^{(i)} =0 \,, \qquad A_1^{(i)} \ne 0\,, \qquad i=1,2\,,
\end{equation}
and

\begin{equation}\label{65}
\Lambda_i = \exp\left\{i {\rm \, arg}([A_1^{(i)}]^2 -4 +4 i A_1^{(i)}  )  \right\}\left(\begin{array}{cc} -1 & 0 \\[2ex] 0 & -1  \end{array} \right)\, \qquad i=1,2\,.
\end{equation}
}

For further details, see \cite{BLM24}.

\end{itemize}


\begin{acknowledgments}

CAB, JTL and LAM thank the Departamento de F{\'\i}sica Te\'orica, At\'omica Y \'Optica at Universidad de Valladolid, and in particular Professors Luis M. Nieto and Manuel Gadella, for the warm hospitality during visits at various stages of this work.

MG thanks the Department of Mathematics and Statistics at the State University of Ponta Grossa and Dr. Jos\'e T. Lunardi for the warm hospitality during a visit. 

LAM thanks the NASA - MN Space Grant Consortium, and Concordia College's Office of Undergraduate Research, Scholarship, and Creative Activity for partial financial support.

{This research was partially supported by the Q-CAYLE project, funded by the European Union-Next Generation UE/MICIU/Plan de Recuperaci\'on, Transformaci\'on y Resiliencia/Junta de Castilla y Le\'on (PRTRC17.11), and also by projects PID2023-148409NB-I00, funded by MICIU/AEI/10. 13039/501100011033I. We also acknowledge the financial support of Castilla y Le\'on  Department of Education  and the FEDER Funds (CLU-2023-1-05).}
\end{acknowledgments}

\bibliography{2RelSingBarriers_v4.bib}


\end{document}